# Towards resilient cities: A hybrid simulation framework for risk mitigation through data-driven decision making

David Carramiñana [a,*], Ana M. Bernardos [a], Juan A. Besada [a], José R. Casar [a]

[a] *Information Processing and Telecommunications Center, Universidad Politécnica de Madrid, 28040 Madrid, Spain*

ARTICLE INFO

ABSTRACT

Providing a comprehensive view of the city operation and offering useful metrics for decision-making is a well-known challenge for urban risk-analysis systems. Existing systems are, in many cases, generalizations of previous domain specific tools/methodologies that may not cover all urban interdependencies and makes it difficult to have homogeneous indicators. In order to overcome this limitation while seeking for effective support to decision makers, this article introduces a novel hybrid simulation framework for risk mitigation. The framework is built on a proposed city concept that considers the urban space as a Complex Adaptive System composed by interconnected Critical Infrastructures. In this concept, a Social System, which models daily patterns and social interactions of the citizens in the Urban Landscape, drives the CIs demand to configure the full city picture. The framework's hybrid design integrates agent-based and network-based modeling by breaking down city agents into system-dependent subagents, to enable both inter and intra-system interaction simulation, respectively. A layered structure of indicators at different aggregation levels is also developed, to ensure that decisions are not only data-driven but also explainable. Therefore, the proposed simulation framework can serve as a DSS tool that allows the quantitative analysis of the impact of threats at different levels. First, system-level metrics can be used to get a broad view on the city resilience. Then, agent-level metrics back those figures and provide better explainability. On implementation, the proposed framework enables component reusability (for eased coding), simulation federation (enabling the integration of existing system-oriented simulators), discrete simulation in accelerated time (for rapid scenario simulation) and decision-oriented visualization (for informed outputs). The system built under the proposed approach facilitates to simulate various risk mitigation strategies for a scenario under analysis, allowing decision-makers to foresee potential outcomes. A case study has been deployed on a framework prototype to demonstrate how the DSS can be used in real-world situations, specifically combining cyber hazards over health and traffic infrastructures. The proposal aims at pushing the boundaries of urban city simulation towards more real, intelligent, and automated frameworks.

## 1. Introduction: decision-making to enhance cities resiliency

Demographic trends over the last century have transformed cities into the cornerstones of modern societies: the world's population has concentrated in cities, turning them into social, economic, technological and innovation poles [1]. Their success is the result of the synergistic socioeconomic interaction between citizens within the city. Due to these interactions, **cities are complex systems composed of multiple inter-reliant domains that include social and economic interactions, governance networks and Critical Infrastructure (CI) systems** [2].

* Corresponding author.
*E-mail address:* d.carraminana@upm.es (D. Carramiñana).

However, this complexity and interdependences make **cities especially vulnerable to hazards whose impact can propagate across different infrastructures**. In recent years, urban societies have experienced the occurrence of risks that have demonstrated the wide-ranging consequences that they entail. For instance, the SaRS-COV-2 pandemic has generated unprecedented disruption in cities [3]. In addition to the obvious cost in human losses, it changed citizens' mobility, work, and consumption habits; with repercussions that are still visible today. In a short period of time, the imposition of confinements meant a shift from public life to the virtual world with the consequent effect on communications and electricity distribution networks. Examples of compounding threats were also experienced with systematic cyberattacks affecting collapsed hospitals [4]. Also, recent conflicts have prompted multiple examples of cascading failures across the electrical grid, water distribution systems, heating systems and mobility systems [5].These examples prove that infrastructures, in general, and within cities in particular, are usually not sufficiently well prepared to cope with existing natural and deliberate threats. As a result, **there exists a city-planning trend to create resilient cities**, which can be understood as cities that are able to absorb, adapt and recover from the effects of a threat without generalized chaos or harm [6].

Enhancing the resiliency of cities requires a comprehensive risk analysis process that can be used to guide city-planning decisions. However, this process is challenging due to the complexity of understanding and modeling cities operation. In fact, to ensure the effectiveness of planning decisions and to make optimal use of available resources, these processes cannot be guided by the intuitions of decision-makers but must be data-driven. In this sense, **technology-aided decision-support systems (DSS) can help in the urban planning process by providing helpful insights and metrics based on real data or reliable simulation models**, contributing to city resilience [7].

As it will be discussed in Section 2, some existing challenges in DSS are the difficulty of providing a comprehensive view of the city operation and offering risk-related useful metrics to the decision makers. To mitigate this gap, Section 3 analyzes the city operation as a system of intelligent and measurable systems. Cities are composed of several interdependent CIs that operate within the city geospatial context and are driven by citizens behaviors. Based on these ideas, a **conceptual model that provides a holistic, formal view of the city** is suggested. This theoretical contribution defines the different critical systems of the city based on a literature review. Additionally, as a novelty, it proposes to consider a social, urban landscape and environment systems to encapsulate the modeling of citizen behaviors and geospatial interactions. Then, the article proposes to identify interdependences between different systems through influence diagrams. In a smart city, the operation of each system can be monitored in real time thanks to IoT devices. A relevant contribution of this article is the identification of how these metrics can be aggregated at different levels (i.e.: entity, system, and city levels) to assess the operation of the city and guide a data-driven decision-making process.

Another relevant feature of modern DSSs is their ability to provide predictability using simulation models. In this sense, the article builds upon the previous conceptual all-encompassing model of the city to propose an **implementable simulation framework consisting in a hybrid approach that integrates agent-based modeling and network-based modeling**. Agent-based modeling (ABM) enables studying the emerging behaviors of the city while network-based modeling allows to easily articulate the interdependences between different entities. Thus, the city is modelled as a set of autonomous agents (i.e.: entities within the city) each composed by a set of subagents and encompassing stochastic elements. Within an agent, each subagent operates in a different network (i.e.: critical system) modeling dependencies among a single system. Then, inter-system dependencies are modelled within the agent by defining the interactions between the different subagents. As a novel characteristic, the proposed hybrid framework eases the modeling of complex interdependences as it can be implemented using object-oriented-programming enabling component reusability. Moreover, this structure also allows to perform simulator federation which facilitates the integration of pre-existing simulators. Also, as a valuable output for decision-makers, the model naturally provides metrics at different aggregation and explainability levels. The proposed simulation framework is detailed in Section 4.

By fusing the conceptual model and the hybrid simulation framework, this article aims to provide a valuable DSS tool that can assist decision-makers in taking informed decisions aimed to enhance city resiliency. The tool allows data-driven decision making and includes explainable elements to compare decisions This practical application is explored in Section 5 where a representative decision scenario explores the interdependences between **the healthcare, ICT, and mobility systems within the city in the presence of compounding threats**. To the authors knowledge, the interactions of these three critical systems have not been previously discussed and modelled in the literature. In this context, and inspired on the Covid-19 situation, the decision-maker must take alternative resiliency-enhancing decisions to mitigate the impact of and a pandemic and concurrent cyber-attacks. The scenario proves how the proposed framework can be integrated into a risk-informed decision-making process and provide useful information to the decision-maker, constituting a powerful DSS.

## 2. Related work

According to [8], threats affecting cities are diverse and can range from natural catastrophes, adverse climatic events or pandemics to financial crises, political instability, hybrid conflicts, terrorism or technological failures (e.g.: cyberattacks, infrastructure failures). In fact, multiple of these threats can materialize simultaneously (compounding threats) or even as a set of interrelated and sequential threats (cascading threats) is also possible [9]. Once a hazardous event occurs, it can have a negative impact (i.e.: a series of consequences) on city systems depending on their vulnerability. Moreover, the occurrence of a threat does not impact only one part of the city. Due to interdependencies, failures spread between systems (i.e.: cascading effects) disrupting the city as a whole and even generating long-term effects. The intersection of a threat, its associated likelihood and its impact on a system can be denoted as a risk

[10].

To implement a resiliency-enhancing city planning process, it is important to assess cities' risks: identifying and analyzing hazards and threats to determine their likelihood and consequences. In some cases, analyzed risks can exceed acceptable risk thresholds and urban planning must take appropriate actions to reduce that risk. In this sense, risk mitigation measures can be taken to eliminate or reduce their impact on the city [11]. When multiple alternative mitigation actions are possible, a risk-informed decision-making process can be used to compare the different decision options based on risk information. Moreover, within the resilient city context, risk mitigation must be understood as a distributed process between different infrastructures that can collaboratively implement actions to benefit the whole city.

As discussed in the introduction, performing risk-informed decision making within the city is a demanding task. Previous studies have already proven that management processes and decision science tools can be used to enhance urban resilience [7]. Thus, one can resort to the use of DSS systems for risk management in cities, together with inference or simulation tools that allow predicting the impact of risks or the benefit of different measures. In order to position the article contributions, this section gathers an analysis of existing tools in the literature. An ample literature review has been previously performed in [12] showing DSS with varied characteristics: considering different geographical levels (e.g.: national level [13] or city level [14]), targeting different audiences and decision horizons (strategic [15] or tactical decisions [16]), providing different views on the city (e.g.: a comprehensive interconnected view [17] or only a subset of systems [18]), encompassing different stages of the decision process (e.g.: only risk assessment [14] or supporting the full decision-making [18]) and generating different types of insights (e.g.: qualitative (e.g.: CARVER2) or quantitative metrics [13]). Some relevant examples of the varied nature of these systems are provided below.

For instance, within the tools reviewed in [12], CARVER2 is a tool developed to analyze risks in critical infrastructure devoted for policy-makers. Thus, it provides a non-technical impact assessment method based on a series of qualitative estimates (e.g.: criticality, possible interdependences) introduced by the decision maker. As a result, the system offers various forms and metrics that allow cross-sectorial analysis, also considering interdependencies.

Alternatively, the Critical Infrastructure Protection Decision Support System (CIPDSS) [13] implements a risk-informed decision making process. It is a decision-support tool that performs a risk assessment process using high-level simulation at a national level. The tool allows to assess the impact of a given uncertain event considering interdependencies. Then, decision makers can compare the effect of different mitigation measures using a proposed set of common aggregated metrics (e.g.: fatalities, economic loss…).

In [18], a DSS deployed to enhance the resilience of strategic critical infrastructure networks at a national level is proposed. The tool specifically focuses on the identification and prediction of hazardous natural events using a variety of data sources (e.g.: weather forecast, seismic data…). It also allows predicting the impact and consequences on critical infrastructures such as roads or the electrical grid considering the interdependencies between networks, and supporting the planning of mitigation strategies through simulation. In [17], the proposed DSS consists in a high-level risk manager where decision-makers must identify threats and provide estimates of their consequences in CIs of interest. Comparison of different decision alternatives through a risk reduction assessment process is possible. However, this process is not guided with simulation but with user-provided information on a set of predefined categories. Likewise, interdependencies between CIs are not generally considered within the proposed model.

Focusing on tools created for cities, an integrated resilience system (aka DSS) is proposed in [14]. As a novelty alongside previous proposals, risk assessment is performed with both models and real data. Thus, the system interacts with CIs to gather real-time information that enables quick incident reporting in emergencies (intelligent monitoring with Artificial Intelligence is considered). Then, this historical information can also be used for multi-hazard risk analysis together with a simulation engine. The simulator can model critical infrastructures (and their interdependencies) to predict the possible the impact, the response and the restoration for a given risk. However, the provided examples use simulation and prediction as an early-warning tool, not for infrastructure decision making.

Authors in [15] also consider the use of massive data for the resilience assessment of critical infrastructures in cities. A methodology is proposed to assess resilience and assist decision-making. First, real-time information is gathered which can be used to evaluate the evolution of resilience using a resiliency curve. Then, a city model is created (digital twin where interdependences might be considered) to be used for decision-making. Finally, different alternative decisions can be evaluated analyzing the resulting changes in the resilience curve through simulation. Authors explore the usage of the tool for short-term decision, the usage of the methodology for long-term planning in cities is not discussed. Likewise, technological details on the DSS tool are not provided.

Real time decision-making in cities is also explored in [16], where a DSS to respond to crisis is proposed. Here, real-time information from CIs and an agent-based simulation model of the city is used to propose decisions and strategies that improve urban resilience. Initially, multiple disruption scenarios are simulated which might be resolved by a series of countermeasures. An urban resilience assessment with multiple KPIs is then made to allow comparing different measures and strategies. The results of these analysis are stored together with the city landscape that arose that countermeasure. Once the disaster situation arises (detected from the received real-time data), the decision-maker is quickly provided with the set of precomputed decision alternatives.

Overall, authors of [12] and [14] conclude that there is a gap in how to approach disruption and disaster situations in cities that would require further investigation to improve existing DSS. First, in many cases the proposed tools are generalizations of previous domain specific (only one or some systems covered) tools/methodologies. However, these generalizations might not be able to cover all interdependencies and other system-specific knowledge within complex systems of systems. Thus, **a comprehensive framework would be preferrable**. In fact, this holistic approach would allow tackling risk mitigation from a resilience point of view via coordination and balancing between multiple networked infrastructures. In that line, authors in [19] also defend that the decision-makers should be presented with a systematic DSS (both in risk assessment and planning) to help them make decisions about which actions, plans and policies to implement to make resilient cities. In that regard, to provide a global view of the city encompassing its complexity, resiliency assessment systems are expected to include the evaluation of all city entities (including all sectors: economic, social, physical, and institutional) and to consider their interconnections.

Authors in [7] argue that there is a need to provide spatial representations of data that helps decision makers to better understand situations. In that sense, GIS techniques can provide spatial context to data helping decision makers to understand the problem and evaluate alternatives in a more comprehensive way. Alternatively, **explainable indicators or advanced dashboarding techniques can also be used to help decision-makers to interpret DSS results**.

Reviews also show that *DSS tools* **are expected to provide predictability via city simulation models that encompass stochastic components**. The models are anticipated to generate qualitative and quantitative metrics to support multicriteria decision making techniques that can help decision makers to evaluate and compare alternative options for risk reduction and response. In that sense, another existing gap is related to the need of providing comparable performance and impact metrics among different sectors (e.g.: electricity grid and healthcare system). Even though the use of aggregated economic or social impact figures and the use of scoring systems has already been explored, additional research along this line would be beneficial.

In order to provide predictability, it is necessary to model the different systems of the city taking into account their complexity and interconnections. Modeling and simulation of city infrastructures has been well studied in literature as discussed in reviews [9] [20, 21],. Traditionally, several methods have been used, as described in Table 1. Particularly, **agent-based modeling and system-dynamics methods are well suited to model CIs** as complex systems, either following a down-top approach in which individual entities and interactions are modelled or following a high-level system view respectively.

Each of the traditional approaches captures a different perspective of interdependent CI (e.g.: geographical relationships can be well explained with a network approach but can be difficult to represent with a systems dynamics approach) and require different resources (in terms of data and computational cost). As a result, **integrated or hybrid approaches** (also included in Table 1) **are also found in literature complementing several techniques to provide a comprehensive view on critical infrastructure operations** [20]. In general, hybrid systems in the literature seem to integrate two modeling approaches. First, network-based modeling is used to organize and model interdependencies between system or entities. Then, a second approach (either, system dynamics, economic-based modeling or ABM) is used to model the operation of each of the nodes within the network. For instance, network-based modeling is used in [22] to organize interdependencies between entities of different systems. Then, a system-dynamic approach is followed to model each entity through a set of state machines with associated differential equations. A gap in this method is that logical and geographical dependencies are not covered. In [23], a generalized framework is described based on the usage of an economic-based approach on top of a multilayer network. Geographical and inter-system dependencies are modelled through the network while the detailed simulation is based on a goods flow model.

Focusing on CI within cities, similar techniques are used due to the importance of infrastructure networks within them. An agent-based approach is used in [24] to model electrical vehicles charging within a smart city. The transportation system and the electrical grid are divided into a set of agents: vehicles, charging stations… Then, the inner operation of each agent is individually modelled along with their interactions. An agent-based method is also explored in [25],[26], with the novelty of introducing disruptions (i.e.: hazardous events) and the possibility of intelligent agents implementing corrective measures to adapt to them. Hybrid approaches are also used here, as in [27] where a multi-layered agent-based model is introduced. In this hybrid proposal, agents are grouped in different layers that represent abstract organizations with specific goals. This allows modeling inter-system and intra-system dependencies at different levels. Overall, even if abstract modeling frameworks are proposed, the particularization of such frameworks in the context of a city is not explored in a comprehensive way.

**Table 1**
Analysis of modeling techniques to represent cities.

| City Modelling Approach | Description | Advantages/Disadvantages |
|---|---|---|
| Empirical techniques | Historical data and expert experience is used to identify and statistically model frequent patterns such as failures, interdependencies | These methods are affected by a reporting bias and cannot provide good estimates for new unknown types of disasters |
| Agent-based modeling | CIs are considered as complex adaptive systems (CAS) that can be modelled following a bottom-up approach. Here, the behavior of the complex system emerges from the interaction of many individual, autonomous, generally simple agents. Agents can represent individuals, organizations, or even abstract entities, depending on the context of the modeled system. | They can capture all types of interdependencies between systems but are difficult to calibrate and validate. |
| System dynamics methods | A top-down approach is used to analyze CIs as CAS. Each system is modelled by a set of variables representing their state that evolve over time by flow rates between systems, usually modelled with differential equations. The dependences between systems are modelled by a series of diagrams that govern causal relationships and product flows. | Stablishing the differential equations and diagrams might be challenging, requires expert-knowledge and cannot analyze component level behavior |
| Economic theory approaches | They analyze CI dependencies through models of economic relationships analyzing the interchange of economic goods between them | Not all systems (e.g.: healthcare) and relationships (e.g.: weather) are observable from an economic view. |
| Network-based methods | Relationships are described by networks/graphs where nodes represent CIs and edges model physical and relational connections. Over these networks, the flow of goods and commodities can be simulated. They can also serve as a basis for point of failures analysis. | They might not be suitable for the representation of isolated CI complexity. |
| Hybrid-based methods | Integrate modeling approaches in order to be able to organize the relationships among systems or entities, while also being able to show their internal dynamics. | They achieve a balance between the representation of micro and macro dynamics, albeit at the expense of elevating the complexity of the design model. |

**Table 2**
Declared Critical Infrastructures by country and proposed city systems. The references cited in this table are [30–34]

| Considered critical infrastructure sectors in each country | | | | | Proposed systems in the city | |
|---|---|---|---|---|---|---|
| **United States [30]** | **Germany [31]** | **Spain [32]** | **United Kingdom [33]** | **European Union [34]** | **Aggregated Sectors** | **Example of entities within sector** |
| Nuclear reactors, materials and waste | - | Nuclear | Civil nuclear | - | Electricity grid | Power plants, power grid, control elements, consumers, prosumers… |
| Energy | Energy | Energy | Energy | Energy | Gas & Oil distribution | Gas and oil pipelines, storage plants, oil distribution trucks, distribution networks, gas stations, consumers… |
| Water and wastewater systems | Water | Water | Water | Wastewater | Water management | Dams, water wells, treatment plants, water pipes, valves, consumers… |
| Dams | | | | Drinking water | | |
| - | Municipal waste disposal | - | - | - | Waste management | Waste trucks, recycling plants, dumpsters, landfills… |
| Communications | Information Technology and Telecommunication | ICT | Communications | Digital infrastructure | ICT | Fixed or mobile communication networks, data centers… |
| Information technology | | | | | | |
| Transportation systems | Transport and traffic | Transportation | Transport | Transport | Transport & mobility | Vehicles, roads, traffic control systems, underground systems… |
| Chemical | - | Chemical | Chemical | - | Food & Goods | Agricultural areas, factories, wholesale market, logistic centers, shopping malls, convenience stores, consumers... |
| Commercial facilities | - | - | - | - | | |
| Critical manufacturing | - | - | - | - | | |
| Food and agriculture | Food | Food | Food | Producing processing and distribution of food | | |
| Financial services | Finance and insurance | Finance | Finance | Financial market | Banking & Finance | ATMs, bank offices, stock exchanges… |
| | | | | Infrastructure Banking | | |
| Healthcare and public health | Health | Health | Health | Health | Healthcare | Hospitals, clinics… |
| Emergency services | - | - | Emergency services | - | Emergency services | Police departments, fire department, public works departments… |
| - | Media and culture | - | - | - | Education, media & culture | Schools, high schools, universities, cultural centers… |
| - | - | Research | - | - | | |
| Government facilities | State and administration | Administration | Government | Public administration | Government | City council, lawmakers, regulations… |
| Defense industrial base | - | - | Defense | - | Not applicable | |
| - | - | Space | Space | Space | Not applicable | |
| | | | | | Social System | Citizens, households… |
| | | | | | Urban Landscape | Facilities, houses, streets… |
| | | | | | Weather | Weather, pollution... |

Taking into account this literature review, the introduction of the hybrid approach for city modeling proposed in this article addresses significant shortcomings in existing methods, particularly the absence of a conceptual model rooted in systems theory that can effectively guide the development of simulation frameworks. Unlike conventional approaches, the proposed hybrid model makes possible simulation federation (or co-simulation), a technique often overlooked in the urban simulation field. This method involves creating a comprehensive simulation model by seamlessly integrating various off-the-shelf system-specific simulators. The innovative aspect lies in the reduction of the modeling effort required, a distinct advantage over traditional methods. This streamlining of the modeling process not only enhances efficiency but also ensures a more realistic representation of diverse systems within a city, leveraging the strengths of domain simulation while ensuring versatility to include specificities. Additionally, the hybrid model is specifically designed to focus on decision-making through layered aggregated indicators, a critical aspect often neglected in prior models. By emphasizing this dimension, the hybrid approach can result into improvements over existing city modeling methods, showing a more comprehensive and efficient tool for urban simulation.

## 3. A conceptual model of cities: the city as a system of systems

Following on the identified need to have a comprehensive approach to understand and deal with the complexity of cities, an urban city model that presents the city as set of interconnected systems is proposed in Section 3.1. Moreover, to enable data-driven decision-making, measurements at different levels are explored to assess cities' performance (Section 3.2). This section lays out the general, theoretical, formal model of cities that is used in next section to define a hybrid simulation framework that supports risk-informed decision-making.

### *3.1. The city structure and operation*

Cities are composed of multiple physical elements (e.g.: parks, roads, transport stops…) and infrastructures (e.g.: electricity, water…) that are used daily by citizens. From an abstract point of view, these elements, including the citizens, can be thought as autonomous entities that make up cities. As previously introduced, these city entities interact with each other. In fact, the high-level city behavior and operation emerges from these low-level interactions. Following on this idea, the city can be understood as a Complex Adaptive System (CAS) [28]. Under this paradigm, cities are dynamic systems in which multiple entities relate to each other over time showing non-linear and adaptative behaviors. Besides, city's entities also exhibit emergent and organized behavior consisting in the formation of larger entities or operations (i.e.: other subsystems) at different scales. These subsystems, in turn, also relate and interact to each other, forming new systems. Overall, and as discussed in [24,29], **the city can be seen as a "system of systems"** where the proper functioning of the whole system (i.e.: the city) emerges and depends on the functioning of a set of interrelated systems (i.e.: each of the essential services), each of them composed of a set of autonomous entities.

Although city systems are also formed in the domain of human interactions and in the governance bodies of cities, infrastructure systems providing basic services (e.g.: energy and water distribution, waste management, transportation, health systems, digital and communication systems…) are essential in the functioning of cities [29]. In general, these infrastructures are known as Critical Infrastructures (CI) which, paraphrasing the definition provided in [30], are those systems or assets that are so vital to the functioning of a city that their incapacity or destruction would have a debilitating impact on security, economy, health or safety. Building upon the previous "system of systems" idea and considering that the operation of the city depends on the correct operation of critical infrastructures, the city can be analyzed **as a system of interrelated critical infrastructures**. Therefore, to build a conceptual city model under this paradigm, two questions must be discussed:

A) Which systems/critical infrastructures make up the city? What entities make up each of the systems?
B) How are these systems interconnected through a series of dependency relationships?

Regarding the first question, there is not a single definition of which infrastructure systems should be considered critical: each country defines those most relevant to its national security. However, an analysis of CI definitions from different countries (as the one depicted in Table 2) can help identify which critical infrastructures are relevant for cities.

The energy sector appears as a critical infrastructure in all the analyzed countries. Specifically, the electricity sector is ubiquitous in modern societies, enabling the operation of all types of processes and systems (e.g., transportation, manufacturing, communications, etc.). Therefore, its interdependencies with the rest of the CIs to which it provides energy must be carefully considered. This sector includes not only the distribution and consumption processes of energy but also its production through different power plants, including nuclear energy. Continuing in the energy sector, gas and oil distribution is also critical for transportation or heating within the city. However, the nature of these distribution networks is different from the electrical networks due to the possibility of storage and non-networked distribution. Hence its modeling as an independent critical system.

Another of the fundamental service networks is water management, including drinking water purification, distribution, and wastewater treatment. Likewise, the waste management system ensures the sanitation of cities through street cleaning and waste removal and processing. The four previous systems require control and communication systems to ensure their stability and real-time operation. Therefore, information and communication technology (ICT) systems also have a critical role within cities, going beyond citizen-to-citizen communications. Similarly, the transportation and mobility system also have strong dependencies with other systems by enabling the distribution of goods and citizens mobility. This system includes both road traffic in the city and collective transport systems such as railroads or bus networks.

Services provision is a fundamental part of the economy of cities. This importance is reflected in the existence of the critical food and goods sector, which groups critical infrastructures in different countries related to the production and distribution of both food and manufactured goods, including commercial facilities, and the provision of final services to citizens. Underlying this, the banking and finance sector enables monetary transactions between different citizens or companies and provides financing and securities.

The SaRS-COV2 pandemic has demonstrated the importance of health infrastructures, which is recognized in the CI repository of all countries. Included in this system are hospitals, primary care clinics, and health centers. Similarly, emergency services such as police, fire departments and essential public workers help to maintain order and safety for people and property. In this way, they ensure a safe ecosystem in which all other infrastructures are deployed. They are also essential for restoring the city's functioning in the event of an emergency or disaster.

Some countries identify the media and research as critical sectors. Although, the education, media and culture sector may seem less relevant in the daily operation of cities, it reports public announcements in emergency events and has long-term impact in the city. In fact, this system is essential to educate future citizens and to build the city of the future through research. Finally, the government sector establishes the conditions under which city activities are carried out through legislation and the promotion of projects. Other sectors such as the defense and space sectors are relevant at the country level, but do not have an immediate impact on the operation of cities and are therefore not considered.

Although those critical infrastructure systems that make up cities have already been identified, it is also necessary to model other aspects of the city. The demand of each of the previous systems arises from the daily patterns and social interactions of the citizens of the city. Therefore, our proposal includes to extend the previously defined systems within the city to also consider a **social system**. This system models the behavior of citizens (i.e.: daily movements and activities), their interactions between them (e.g.: households, social gatherings) and with other systems (e.g.: electricity consumption). Similarly, CI systems are also modified and affected by the geospatial context they are deployed on. This includes the physical form of the city (e.g., its streets, parks) and environmental elements (e.g., weather, pollution). Therefore, to complete the formal model that make up the city as a system of systems, our proposal also includes a system to model the urban landscape and another to consider the environment. As a summary, Table 2 also depicts the proposed city systems and a set of entities that are included/participate in each of the systems.

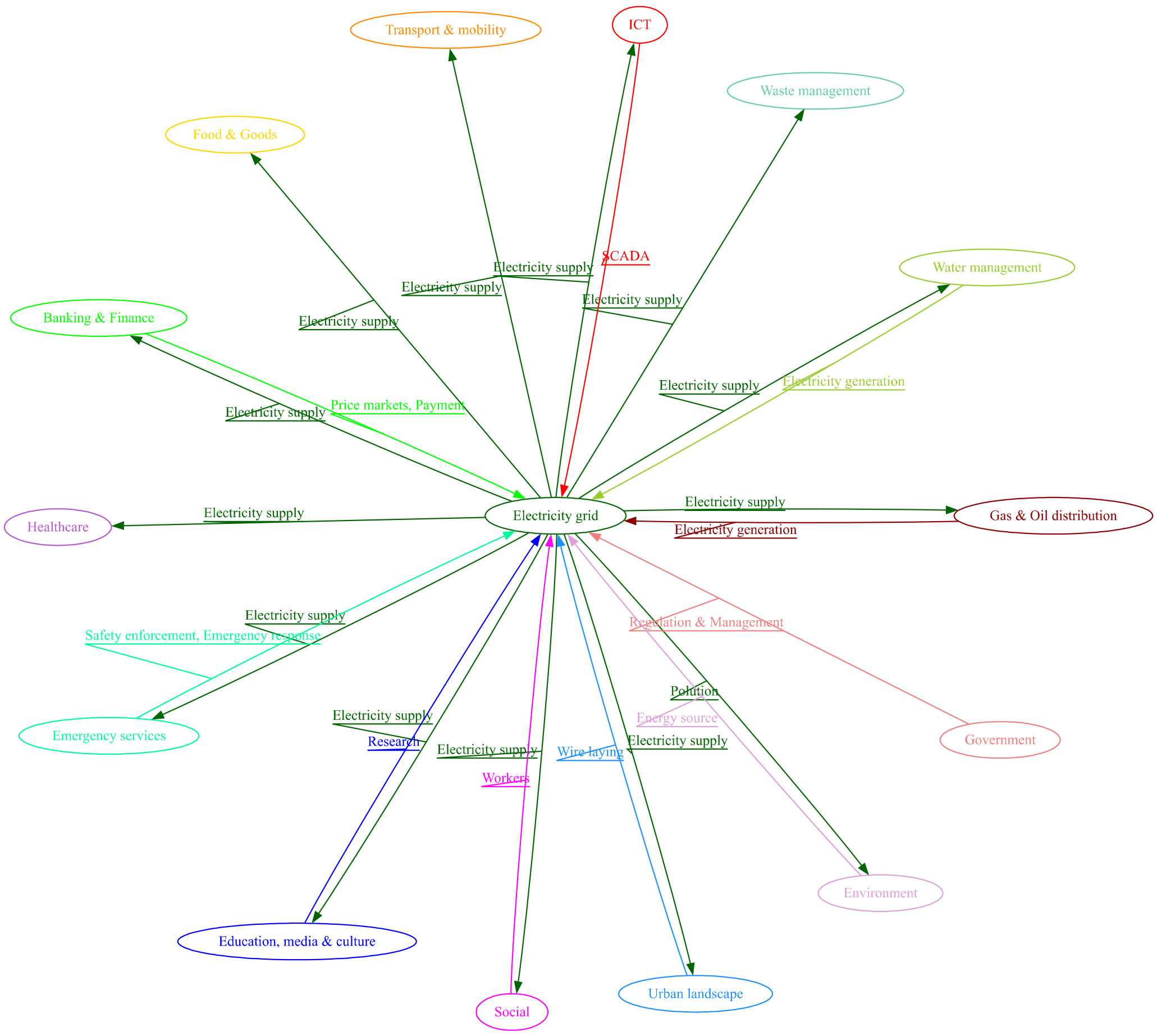

**Fig. 1.** Example of influence diagram representing interdependencies of the electricity grid system.
The outgoing arrows from each system indicate that the source system provides a service or supplies a product to the target system as indicated in the arrow tag.

**Table 3**
Performance metrics for each of the identified systems within the city.

| City Systems | Performance Metrics | Examples of open data sources for performance analysis |
| --- | --- | --- |
| **Electricity grid** | Consumption statistics.<br>Production per source.<br>Network stability statistics (frequency…).<br>Detected outages.<br>Power plant availability<br>Price | ENTSOE Transparency Platform[1]: Real-time demand, generation information at country level in Europe |
| **Gas & Oil distribution** | Consumption statistics.<br>Network nodes (pipes, valves…) status (pressure, flow)<br>Available gas and petrol storage<br>Price | ENAGAS open data[2]: Gas demand and storage in Spain |
| **Water management** | Consumption statistics.<br>Network nodes (pipes, valves…) status (pressure, flow)<br>Network expected maintenance works.<br>Water quality<br>Water reserves | Embalses.net [3]: Water reserves in Spain.<br>Canal Isabel II [4]– Real time water disruptions in Madrid (Spain) |
| **Waste management** | Weight of recollected waste.<br>Weight of treated waste. | Ayto. Madrid Open Data [5] – Aggregated waste statistics in Madrid (Spain) |
| **ICT** | Servers and services status.<br>Communication network status and coverage.<br>Number of detected cyberattacks.<br>Unavailability time. | Kaspersky Cyber Map [6]– Real time detected cyberattacks |
| **Transport & mobility** | Maps & Schedules.<br>Traffic intensity.<br>Traffic accidents.<br>Traffic and transport delays.<br>Served passengers statistics. | EMT Open Data [7] – Real time bus network status in Madrid (Spain) |
| | | Ayto Madrid Open Data[8] - Real time traffic lights status in Madrid (Spain) |
| **Food & Goods** | Food production and stocks statistics.<br>Food and goods prices and demand. | INE [9] – Monthly price statistics in Spain |
| **Banking & Finance** | Stock Exchanges prices.<br>Cash distribution network status. | Yahoo Finance [10] – Worlwide, realtime stocks information |
| **Healthcare** | Hospitals capacity and occupancy.<br>Diseases prevalence.<br>Death rate. | CyL Open data[11] – Hospital occupancy in a region of Spain |
| **Emergency services** | Mean time to intervention.<br>Number and distribution of interventions by type (police, firefighting, ambulance). | Ayto Madrid Open Data [12] – Daily interventions of firefighting service in Madrid (Spain) |
| **Education, media & culture** | Education level.<br>Media outlets share. | Ayto Madrid Open Data [13]- Education level by district in Madrid (Spain) |
| **Government** | Elections information.<br>Current regulations. | BOE [14]– Regulations and laws in Spain |
| **Social** | Population statistics. | Ayto Madrid Open Data [15]– Social indicators by district in Madrid (Spain) |
| **Urban landscape** | Infrastructures maps.<br>Urban illumination status.<br>Expected road works.<br>Expected events. | OpenStreetMap – Worldwide geographical information.<br>Ayto Madrid Open Data [16] – Real time events and urban disruptions in Madrid (Spain) |
| **Environment** | Weather status.<br>Pollution levels. | Ayto Madrid Open Data [17] – Real time weather and pollution levels in Madrid (Spain) |

[1] https://transparency.entsoe.eu/dashboard/show
[2] https://www.enagas.es/es/gestion-tecnica-sistema/energy-data/
[3] https://www.embalses.net/
[4] https://oficinavirtual.canaldeisabelsegunda.es/gestiones-on-line/incidencias-en-el-suministro
[5] https://visualizadatos.madrid.es/pages/residuos-recogidas
[6] https://cybermap.kaspersky.com/es
[7] https://apidocs.emtmadrid.es/

Additionally, as already indicated, the relationships and interdependencies among city systems should also be considered in the conceptual city model. Some of these dependencies have already been discussed in the previous description of the identified systems. In a concrete way, **the operation of a city (or of a subset of systems within the city) can be represented using influence diagrams** where each city system is denoted as a node, and relationships and dependences among systems are depicted using labelled arrows. The generation of these diagrams should be based on a systematic analysis in which dependencies are identified system by system. Identification of dependencies can be supported on expert knowledge of infrastructure operation or statistical analysis of real data.

As an example of this process, an expert-based analysis has been performed on the Electricity Grid dependencies resulting in Fig. 1. Within the influence diagram (which only considers first order dependencies of the electricity grid system), the outgoing arrows from each system indicate that the source system provides a service or supplies a product to the target system as indicated in the arrow tag. For instance, the outgoing arrows labeled as “Electricity Supply” indicate that most systems rely on the electricity grid to consume electricity. Conversely, the incoming arrows to the Electricity Grid system indicate that it relies on the ICT system for control and supervision (SCADA), on the Gas & Oil distribution system to obtain fossil energy sources for energy production, on the Banking & Finance systems to negotiate and set the energy prices…

If the previous analysis is repeated for all city systems, it is possible to generate a comprehensive influence diagram by aggregating the different partial diagrams. For clarity purposes, the result of this complete process has been deferred to Annex 1 where the dependencies between all the city systems are shown. The purpose of such an influence diagram is twofold. On the one hand, it allows to clearly identify which relationships must be considered when modeling each system. Then, it also allows a decision-maker to analyze between which cascading effects (first-order, second order… interdependences between systems) may occur as a consequence of a risk. However, as a result of this analysis, it is also possible to conclude that, if no additional structure is set up to organize the dependencies homogeneously, modeling them is a challenging task.

Until now, the proposed model has viewed the city as a self-contained reality without considering the relationships with other cities, national powers… that are relevant to the dynamics of the city. These relationships can also be explored within the proposed conceptual model by including within each of the city systems "boundary entities" that abstractly model the inputs and outputs of the city with the rest of the external entities. For example, a boundary entity within the water management system may serve to model the water reserves available to the city that are managed by a higher-level organization (e.g., hydrographic confederation).

### *3.2. Evaluating and measuring the city operation*

To understand and analyze the behavior of a city, it is necessary to define a set of metrics that describe the level of performance of its different processes and entities. A quantitative city’s operation assessment becomes even more important when, in the context of data-driven resilience enhancement, it is necessary to analyze the impact of risks or to compare mitigation strategies. However, the heterogeneity and complexity of cities makes it difficult to define a single set of explainable metrics that address the main drivers of city performance. In fact, and following the previous formal framework, **these metrics can be defined at different levels of granularity**:

- **Entity level:** very detailed, domain-specific metrics that refer to the attributes of each of the entities of a city. For example, the load level of a tank, the occupancy of a train, the electricity consumption of a house…
- **CI system level.** Ideally, it would be a common, comparable metric that would allow comparing the level of service of each of the CI systems. However, this is usually not possible, and domain-specific aggregated KPIs (Key Performance Indicator) for each of the CI systems are usually used. Some examples would be demand evolution, demand coverage or response time to incidents, particularized for each system.
- **City level.** In practice, these metrics that provide an overall view of the city usually refer to figures that measure the aggregated economic or social impact of an event or measure. For example, changes in a city’s Gross Domestic Product, loss of human lives, variation in the number of citizens… Another possible option is to use a scoring system (i.e.: combinations with different weights) to group heterogeneous metrics obtaining a single KPI for the whole city that considers the impact of the different CI systems [12].

In this regard, Table 3 proposes a set of metrics at the entity or system level for each of the identified systems in the formal city model. In addition, for each of these levels, other filtering and aggregation levels can be applied concurrently, such as: time interval, spatial distribution, attributes of the entities/systems analyzed that give rise to different classifications.

The preference for one level of granularity or another depends on the intended purpose of the metric and the target audience. For example, if they are addressed to the operator of a specific CI system, the operator will require detailed metrics for each of the entities it is composed of. These metrics can be used to make tactical decisions that improve its operation, or to generate targeted alerts on specific elements of the infrastructure that can prompt quick reactive actions. In contrast, a municipal decision-maker (e.g., a politician) requires a global view of the city’s performance with metrics aggregated at the city level or for each of the IC systems. With this holistic view and the comparison of different what-if scenarios; strategic, long-term decisions such as the implementation of preventive measures can be taken. **Ideally, a decision-support system, should be flexible, providing both levels of granularity to suit the needs of the target user.**

A typical challenge in assessing the performance of cities is obtaining the data from which the above metrics are constructed. In this sense, another important pillar of modern cities is their connectivity, monitoring, and intelligence. In recent years, Internet-connected everyday objects and systems have become popular. This paradigm, known as the Internet of Things (IoT), enables more efficient and intelligent services. Particularly, in urban settlements, smart cities have favored the use of sensing and communication technologies in multiple city infrastructures, such as public services, water systems, energy generation, transportation… [35] As a result, the information available on the functioning of cities and critical infrastructure has multiplied, both in real time and in historic datasets [36]. This connectivity and monitoring favor the control and adaptation capacity of the infrastructure systems themselves. Therefore, **modern smart cities can be considered as a system of intelligent and measurable systems**.

The value of this data to analyze and detect risks have been previously recognized in literature as discussed in [14] [15,16],. Particularly, open data initiatives make publicly available useful government-generated information which usually can be ingested programmatically and in a curated format. In fact, multiple open data sources can be fused to provide valuable graphical representations or dashboards [37]. Moreover, the use of cross-domain data can improve the understanding of urban systems creating a comprehensive view that links multiple CI subsystems. In that regard, modern data analysis techniques such as machine learning or artificial intelligence can be used to generate knowledge and recognize systems interdependencies. As an example, [38] explores some use cases where multiple datasets are combined using statistical methods to predict wastewater pipe blockages or parking occupancy. To further prove the usefulness of open-data, Table 3 identifies some valuable open data sources that can be used as a starting-point to derive the city performance metrics.

## 4. A simulation framework for risk-informed decision-making in cities

In the previous section, a conceptual model has been proposed by considering the city comprehensively as a set of interconnected CI systems that are monitored and can be evaluated by a set of metrics. That abstract representation of the city is the first step towards an actionable simulation model that can represent various scenarios and is useful in the context of data-driven decision making. In this sense, Section 4.1 describes a hybrid simulation framework that links ABM and network-based simulation to analyze each CI and their dependencies in the context of the cities' formal model. Then, Section 4.2 points how the proposed simulation framework enables risk-analysis and data-driven decision and describes some implementation details that allow model reusability and simulator federation.

### *4.1. A hybrid simulation framework of the city*

Of the various approaches described for simulating CI in Section 2, a hybrid framework based on agent-based modeling (ABM) [39] and network-based modeling is proposed in this article. ABM fits well with the paradigm of cities as CAS: each of the agents models one of the city functional entities (e.g.: a citizen, a hospital, a power plant…) and the overall behavior of the city emerges from its adaptive, autonomous behavior. In general, ABM agents can be heterogeneous and governed by simple rules that can consider stochastic components. As the city agents interact at different operational levels, different systems are involved (in practice, understood as an aggregation of agents cooperating to fulfill a common goal or provide a common service). Therefore, the city can be modeled as a set of individual autonomous agents that form the different systems identified in Section 3.1:

$$\mathrm{Agents} = \{a_1, a_2 \ldots, a_i, \ldots, a_N\} = \text{autonomous agents in the city model} \tag{1}$$

$$\mathrm{Systems} = \{\mathrm{Healthcare},\ \mathrm{ElectricalGrid},\ \mathrm{WaterDistribution},\ \mathrm{ICT}, \ldots, \text{Generic System}\} = \text{systems in the city model} \tag{2}$$

Modeling the interactions and interdependences between systems can be a challenging task, as demonstrated by the influence diagrams in Fig. 1 and Annex 1. In fact, agents might be part of multiple of the previously identified systems, interacting with other agents within those systems, as shown in the lower half of Fig. 2. For instance, a hospital (modeled as an agent) is part of the healthcare system (e.g.: exchanging patients with other hospitals) and that main function depends on the adequate interrelationships with other systems, which will be functional parts of the same hospital agent: the electrical grid (e.g.: consuming electricity from a distributor), water distribution networks, ICT network (it exchanges information with other hospitals over its own ICT infrastructure)… Each of the agents within the dependency systems, on its own, also depend on interactions with agents of the same systems and possibly other systems. Thus, **the proposed model must reflect actions that occur at two different levels: 1) between different systems (inter-system dependencies) and 2) actions across a single system (intra-system dependencies)**. Network modeling is here used to organize these relationships in a simple way, thus the hybrid approach in the proposed simulation framework.

A **multiplexed-network approach is used to clearly identify and model each of these dependencies while following an agent-based model.** It consists in imposing a multilayer structure that limits agents' interactions in two different planes: the agent dimension and the system layer dimension. As depicted in the upper part of Fig. 2, agents can be considered as being composed of a set of subagents each of them operating in an individual city system while maintaining internal causal and operational relationships among them to model the whole agent behavior. In the proposed implementation, these sub-agents can be standardized to fulfill different roles within a system, allowing to build modular agents by aggregation of standard sub-agents. Individual subagents can be instantiated by providing different parameterization values to each modular agent.

$$\mathrm{agent} \equiv \left\{\mathrm{subagent}_{\mathrm{healthcare}}, \mathrm{subagent}_{\mathrm{electricalgrid}}, \mathrm{subagent}_{\ \mathrm{waternetwork}}, \mathrm{subagent}_{\mathrm{ICT}}, \ldots, \mathrm{subagent}_{\mathrm{genericsystem}}\right\} \tag{3}$$

As depicted in Fig. 2, the inter-system dependencies can be explained by the interactions among sub-agents across the agent dimension. Then, the behavior of a single, complete system (e.g.: the healthcare system) can be explained by the interactions of all the corresponding subagents (e.g.: all the hospitals subagents) across all full city agents that form part of that system (i.e.: not all agents are required to participate in all systems). Therefore, interactions of sub-agents across the layer dimension explains the intra-system dependencies, each of the layers/networks corresponding of the identified city systems.

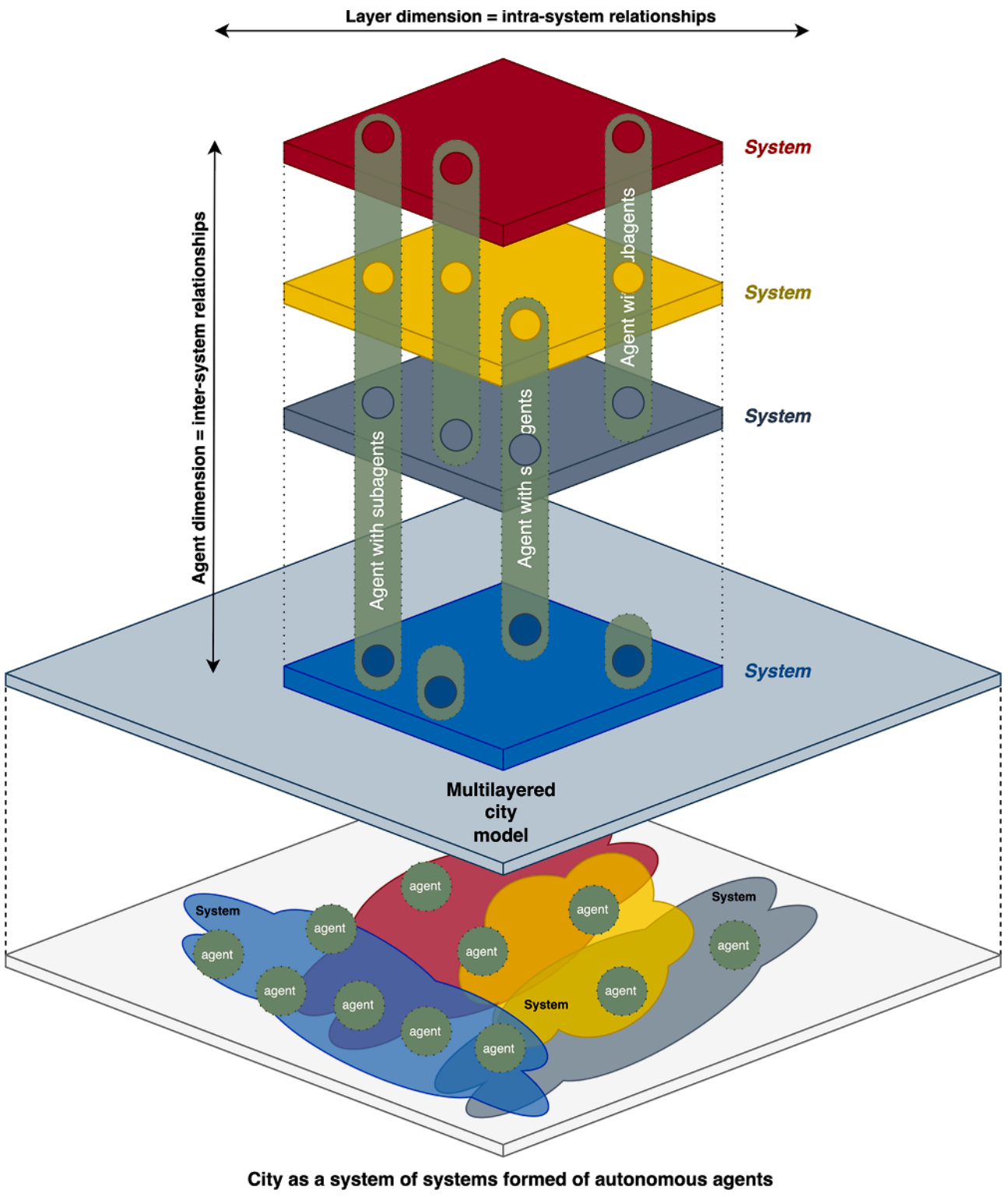

**Fig. 2.** The city as a set of agents organized in systems (down). Multi-layered abstraction to organize relationships between agents and systems (up).

These interactions govern the evolution over time of each subagent internal state (Fig. 3). Multiple tools can be used to define a subagent's state such as finite state machines, status flags or a set of continuous dynamic magnitudes. Interactions can be modelled as well-defined, explainable rules which might consist of a series of conditions, mathematical expressions, or stochastic processes. Rules are expected to be parameterizable at an individual level so that different agent behaviors can be configured. Interaction rules can be decomposed in three different rulesets. First, subagent Internal Rules, (IR) govern the internal operational trends of the subagent that can be explained in an isolated way (do not depend on other subagents). Then, System Interaction rules (SR) represent relationships with other subagents at the same system level. Thus, these interactions explain the network effects within a system, considering how that system subagents affect each other (intra-system dependencies). Finally, an agent encapsulates multiple subagents within a single functional entity. At this level, Agent Interaction rules, AR, explain the relationships between the different subagents of a single agent (intersystem dependencies). That is, they contemplate how other systems affect the behavior of the subagent. To ease the understanding and modeling process, intra-agent dependencies (which will be then made explicit in the AR of each subagent) can be described with an influence diagram mapping dependencies between subagents.

Continuing with the hospital example, the associated healthcare subagent state could consist in its bed occupancy, available healthcare providers or medical stockage. Its IR would model processes such as Emergency Room (ER) operation, wards operation, evolution of healthcare quality. Then, additional subagents would be required to model the hospital's water or electricity consumption would. The provision of those services would be contingent on the interactions modelled with the SR across the corresponding systems. These interactions can affect the healthcare service and vice versa, as modelled by AR. For instance, an increase in patients can result in additional electricity demand; conversely, a blackout can decrease healthcare quality. Overall, the subagent state evolves over time because of the concurrent application of the three set of rules, each considering the state of a given group or interconnected subagents.

$$\begin{aligned}&\text{subagent state}|_{t} = \\ &(\text{AR} \circ \text{SR} \circ \text{IR})\left(\text{parameters, internal state}|_{t-1}, \text{state of subagents within system}|_{t-1}, \text{state of subagents in agent}|_{t-1}\right)\end{aligned} \tag{4}$$

As already discussed, within the scope of a smart city, its entities performance evolution can be monitored over time. For each subagent a subset of metrics can be extracted from its state. This extraction is modelled by an observability function (OB) that allows simulating entity-level metrics for each subagent.

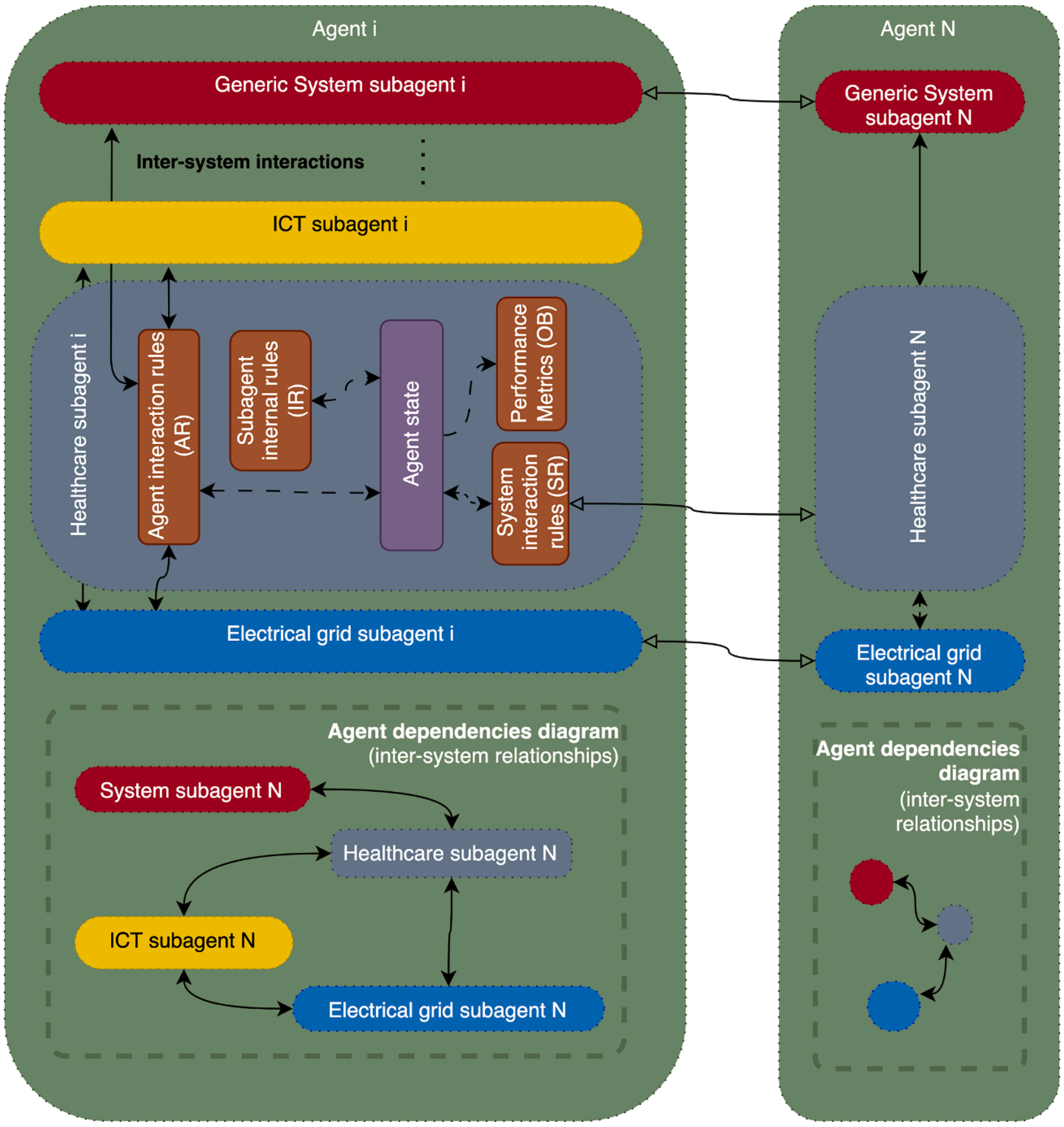

**Fig. 3.** Agent as a set of sub-agents modeling inter-system dependencies.

$$\text{subagent metrics}|_t = \text{OB}\left(\text{subagent state}|_t\right) \tag{5}$$

Summing up, a subagent is composed of a set of parameters, its state, the interaction rules, and its observability model:

$$\text{subagent} \;\; = (\text{parameterization}, \text{state}, \text{IR}, \text{SR}, \text{AR}, \text{OB}) \tag{6}$$

Abstracting to the system level, each system in S will be composed of a set of subagents operating in that system (layer), as shown in Fig. 4. Relationships between subagents (intra-system dependencies) can be easily displayed as a network or dependency diagram where the nodes are the system subagents and edges represent functional or causal relationships between them. On its own, each relationship is modelled by the corresponding SR defined at subagent level.

$$\text{Generic system} \equiv \left\{a_1^{\text{SYSTEM}}, a_2^{\text{SYSTEM}}, a_3^{\text{SYSTEM}}, \ldots, a_N^{\text{SYSTEM}}\right\} \tag{7}$$

Systems performance can also be quantified and monitored within the model. Domain-specific metrics can be obtained by aggregating the metrics of the subagents operating at that level. Specific functions, named Aggregated Observability (AOB), can be used to model the construction of such metrics. Alternatively, a comparable, generic measurement between different systems can also be obtained by quantifying the degree of degradation suffered with respect to the nominal operation (i.e.: correct, without failures) of each system. For this purpose, a metric called **Service Level (SL) is defined. It can be understood as a quantitative, aggregated, system-level metric that is derived from operational metrics of individual modelled agents**. Although operational, domain-specific information might be lost, SL degradations can be still tracked back to individual subagents. If required, city-level metrics could be obtained by using a score function over the SL of all city systems. Therefore, for each city system under consideration, the following can be defined:

$$\text{Metrics}_{\text{genericsystem}}|_t = \text{AOB}_{\text{genericsystem}}\left(\text{state of subagents in system}|_t\right) \tag{8}$$

$$\text{SL}_{\text{genericsystem}}|_t \in [0, 1] = \text{SA}_{\text{genericsystem}}\left(\text{state of subagents in system}|_t\right) \tag{9}$$

Where SA (System Aggregation) is a system-dependent, modeler-defined, monotonically increasing function that allows representing the state of an overall infrastructure system from the observable state of each of its subagents. It is expected that the service level will evolve over time with a SL=0 representing a complete degradation of services and SL=1 representing a full level of service. It is the modeller job to design the *SA* function so that the resulting Service Level accurately represents the operational status of the whole service.

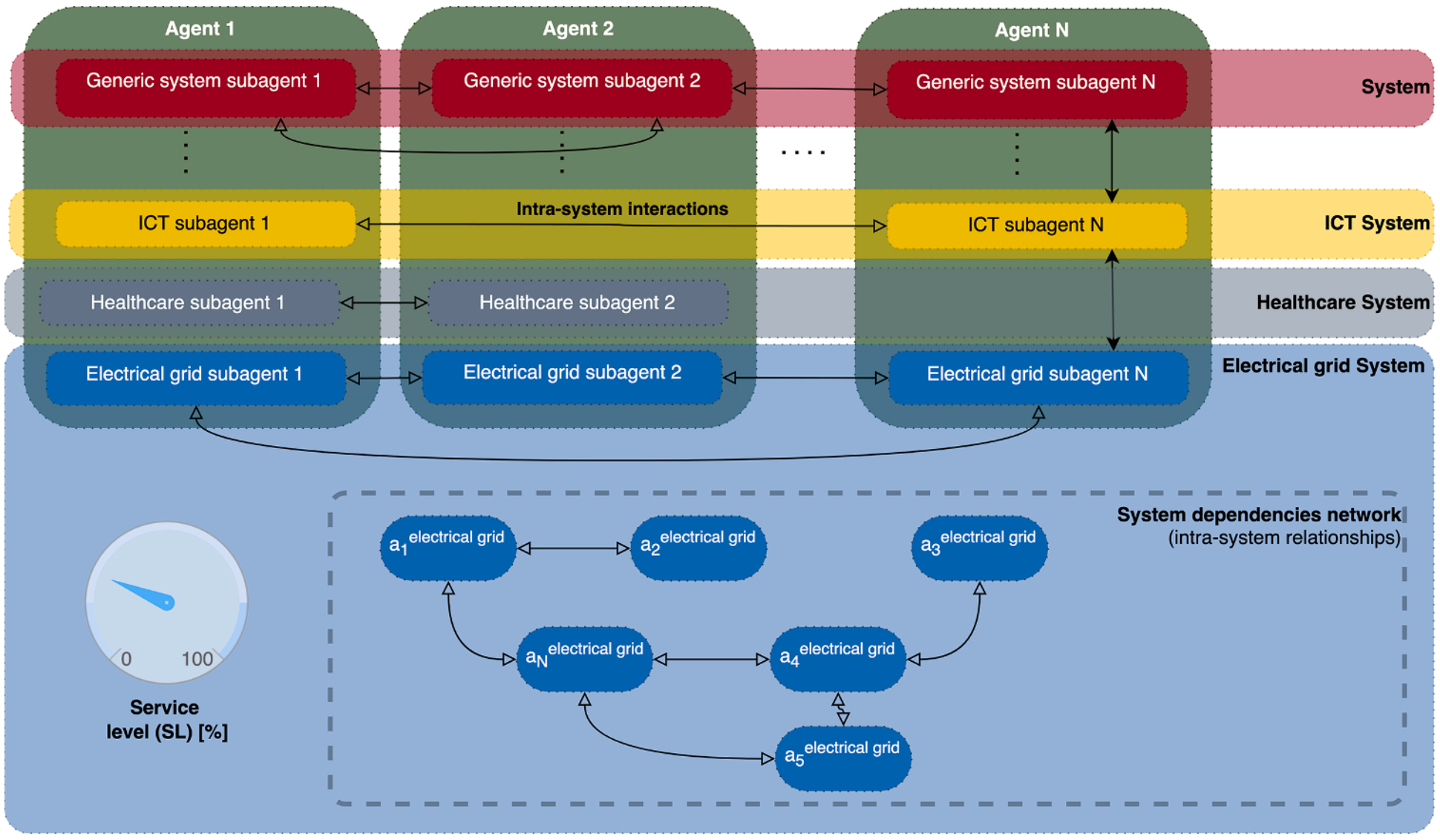

**Fig. 4.** Proposed multi-layered agent-based model.

## *4.2. Adapting the simulation framework to enable risk-informed decision-making*

The goal of the proposed comprehensive simulation framework in the article's scope is to be able to assess the impact of hazardous events on the operation of city and to compare the effect of different risk-mitigation alternatives following a data-driven methodology. In this sense, it is essential to implement the previous simulation framework considering the ease of use from the decision-maker point of view. Therefore, it is desired to reduce the modeling work required to address each decision problem. It is also required to promptly provide results for a broad time horizon so that multiple risk scenarios can be analyzed over the medium to long term. Finally, the simulation results must be easily interpretable by the decision-maker. This subsection discusses how these desiderata can be naturally obtained from the granular nature of Agent-Based Modelling.

Regarding the ease-of-modeling, considering reusability within the implementation allows to leverage previous modeling work by partially reusing it for other problems. To implement this reusable approach, a modular design based on multiple inheritance **object-oriented programming** (OOP) is used, as shown in Fig. 5. First, for each city system under analysis, a series of programming classes are defined and implemented. Each class corresponds to a different subagent within the system, individually implementing the behavior and interactions of an entity of the subsystem based on a state and a set of well-defined rules (the previously defined IR and SR rules along with the observability function). These subagents, within each subsystem, exchange events with each other that model the relevant intra-system dependencies. Then, inheritance is used to build new classes that aggregate the behavior of subagents within their source sub-system while adding an inter-system behavior (AR function). These classes refer to the agents that represent complex entities within the city simultaneously operating in multiple sub-systems. The OOP structure clearly encapsulates sub-agents' behaviors in individual classes, enabling to model each sub-agent at different detail levels and easily swapping different sub-agent models. In fact, it is possible to fully simulate each-subsystem independently. Then, complex systems are not modelled from scratch but by aggregation of the pre-existing modular subagents forming agents. This modular programmatic approach intuitively arises from the agent-based analysis and it enables code reusing. Moreover, each class is parameterizable so that differently configured instances of each subagent and agents can be instantiated in different simulation scenarios.

Another advantage of the proposed modular OOP approach is that it also **allows substituting the modeling of a subsystem by an external simulator**. The integration with external simulators lessens further the modeling and validation effort as well-tested, domain-specific simulators are broadly available. If the external simulator exposes a programmatic or networked API, which is usually available, the ABM simulator can integrate it. To do so, "meta"-subagents are created as the reflection of an external simulation entity within the ABM simulation. These agents oversee controlling and retrieving data from the external simulation. Therefore, they mirror external entities state so that it can be used within the internal simulation. Moreover, they also allow interactions with internal sub-agents that are abstracted from the integration of the external system. Thus, the proposed simulation framework infrastructure is ready to federate multiple simulators, i.e.: to conduct a distributed and coordinated simulation. Even if usually possible, the extent of the integration is dependent on the capabilities of the external simulator. Overall, this feature enables the integration of advanced, complete, and stable simulation capabilities for different city systems. Furthermore, returning to the idea of simulating entities external to the cities, simulation federation would also allow to simulate in parallel multiple cities interrelated to each other by means

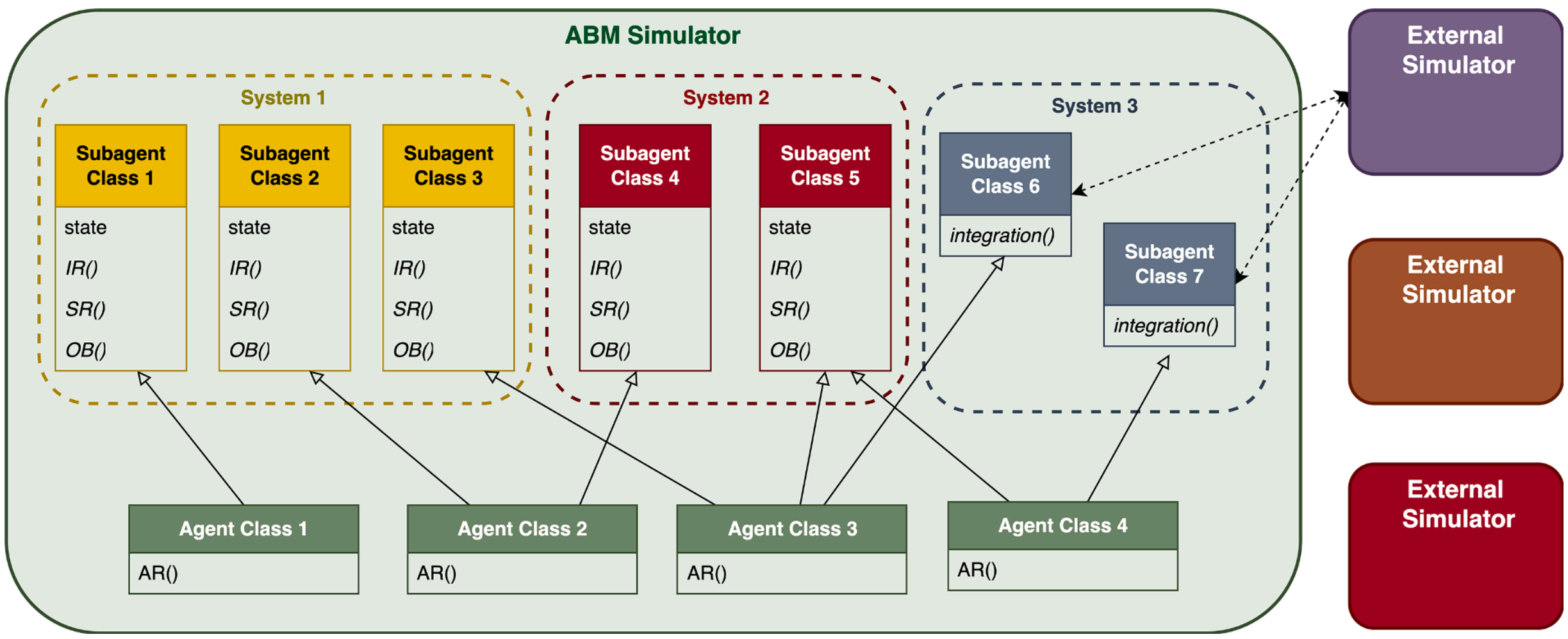

**Fig. 5.** Modular class design for agents and sub-agents based on multiple inheritance.

of the aforementioned boundary entities/agents.

Summing up, the proposed hybrid (i.e.: ABM-network modeling) framework eases the modeling process, as dependencies among systems are included as reasonably well-known, easily explainable, and implementable relationships between subagents. Those relationships can be explained as simple parameterizable rules and mathematical formulas that can be common and standardized across subagents representing a given function within the system. Then, each agent can be properly calibrated to instantiate the actor it represents. Moreover, once a high-level modeling framework (i.e.: types of agents, considered relationships…) has been established, each agent (or subagent) within the simulation can be modelled with a different level of detail. In fact, the model can also be easily extended with the inclusion of new agents of the same or different nature or performing simulators federation.

Another of the desired features to enable risk-analysis is the ability to quickly simulate several scenarios with different risks and implementable decision alternatives. Therefore, the simulation framework uses **discrete simulation in accelerated time**. This type of timing does not introduce dependencies with real time, which allows to quickly generate results for different scenarios using a discretized in simulation time interval. Besides, the framework is configurable in the sense that each agent can be programmatically implemented taking advantage of code reusing between sub-agents or using integration of external simulators. Moreover, each agent depends on a set of parameters that may yield to different simulation scenarios. The configuration capabilities range from the agents to be simulated per system or the dependencies between them. Additionally, **the simulation of contingencies or threats must be considered in the framework**. A hazardous event, HE, can be defined as the occurrence at given time stamps of hazards or threats. Each of them could be parameterized as a variation of the nominal parameters of the affected subagents. Then, because of the described operation, the change of these parameters may have consequences in other subagents due to the interdependencies between them.

$$\text{HE} = \left(\text{hazard}_1|_{t=T}, \text{threat}_2|_{t=T}, \text{threat}_2|_{t=T+2}\cdots\right) \tag{10}$$

These contingencies can be manually defined to occur at a certain time step or use stochastic processes.

Finally, to enable the data-driven approach, the output of the simulation framework is provided through *time series:* the observability function of each subagent and system is used to extract a set of time series with relevant metrics for each agent/system. In this sense, the granularity of the model also allows the generation of metrics that monitor the city for the different levels discussed in the previous section, which helps the decision-maker in understanding the underlying trends. Thus, the framework supports quantitative predictive capabilities to analyze risks. Additionally, it is also possible to compare with the output data different "what-if" alternative scenarios that might also include the existence of non-nominal conditions (hazardous events, contingencies).

These implementation characteristics (i.e.: reusability, simulator federation, accelerated-time simulation, simulation of hazardous event, granular output timeseries) adapt the simulation framework to its use as a decision support tool. For example, the absence of the simulator federation would make it necessary to model complex systems from scratch, increasing the workload required to implement each decision problem in the simulator. Similarly, the lack of simulation in accelerated time would make it impossible to analyze scenarios in the long term because the computational times would be unmanageable. Ultimately, as it will be shown in next section, these features make it easier to carry out simulations that assess the impact of risks and to analyze different mitigation strategies implementing an optimization by simulation approach.

## 5. Validation: integrating the hybrid simulation framework in a risk-informed decision-making process

Once the hybrid simulation framework has been defined, its usefulness as a DSS within a risk-oriented, data-driven decision process must be proven. In this sense, Fig. 6 **envisions a risk analysis and decision methodology that integrates the simulation framework within it.** First, an initial stage allows defining the scope of the problem to be solved: identifying relevant risks to analyze and which city systems (from those identified in Section 3) should be considered. Then, in Stage 2, a modeler would be charged of generating an actionable simulation model of the considered systems as an implementation of the comprehensive hybrid simulation framework in Section 4. This simulation capability enables performing a data-driven risk-analysis in Stage 3 using metrics at different aggregation levels. Afterwards, in Stage 4, the risks are evaluated against acceptable thresholds using the obtained quantitative data. Possible mitigation alternatives are proposed and evaluated with the model to estimate the value of each one in terms of risk reduction. Finally, in Stage 5, the decision maker has a clear picture of the existing risks and a set of assessed alternatives to mitigate them. A comparison between them is possible using the numerical insights provided by the proposed simulation framework to reach a final risk-informed decision to enhance city resiliency.

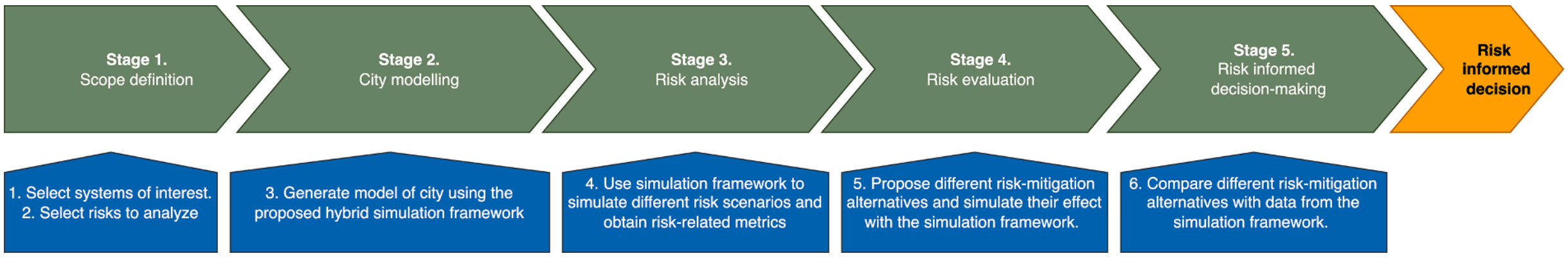

**Fig. 6.** Typical steps in a risk-informed decision-making process.

As a validation, the previous methodology is showcased in a representative use case scenario in which a policymaker wants to assess and improve the resiliency of a given city. Throughout the section it will be showed:

a) How the comprehensive conceptual model of the city proposed in Section 2 provides a solid underlying abstraction to represent and understand the operation of the city.
b) How the hybrid simulation-framework eases the modeling process by organizing entities relationships in different layers. Particularly, a proof-of-concept implementation will be described with a subset of the city systems.
c) How the same framework allows federating other simulators to ease the modeling process.
d) How it also enables a data-driven risk analysis and decision-making process providing quantitative insights at different levels and proving that it constitutes a valuable DSS.

### 5.1. Stage 1. Description of the decision problem

As depicted in the proposed methodology, a description of the problem must be generated first. Although multiple city systems could be contemplated under the conceptual model, the decision-maker is interested in the resilience of the ICT system, the healthcare system and the mobility system of a fictional city. The city under analysis is composed of two geographically separated districts: the city district and the outskirts district.

The ICT infrastructure of the city follows a hierarchical graph in which each district has its own cyber-infrastructure that depends on a centralized city-level node. Regarding healthcare, each district is served by a hospital which is well-dimensioned for the population base demand. Each hospital requires connectivity with the district ICT infrastructure to operate. However, both hospitals operate in a collaborative way to be more resilient. Finally, mobility is simulated over a representation of the city's urban landscape. Traffic lights controlling the vehicles flow in each district are also managed by separate ICT controllers connected to the district ICT node.

Over this scenario, the policymaker wants to assess the impact of a compounding hazardous event (HE) composed of:

- A cyberattack on the ICT infrastructure of the center district, at day 20 of the simulation.
- The simultaneous spread of an airborne infectious disease over the whole city population, from the start of the simulation.

### 5.2. Stage 2. Implementation of a proof-of-concept simulation framework

Once the problem scope has been defined, a city model must be generated in stage 2 to serve as a DSS. In this sense, a prototype of the simulation framework has been developed using Mesa [40], a ABM framework for Python. The implemented simulator only considers a subset of the cities' conceptual systems as required by the problem scope. These systems are (shown in Fig. 7): ICT system, healthcare system, mobility, and transportation system; and two additional supporting systems: the social system, and the urban landscape system. To do so, it integrates these authors' previous works [41] and [42] using the framework proposed in Section 4. Moreover, the capability of integrating external simulators is also demonstrated by including a well-known external simulator (SUMO [43]) to model traffic within the city.

Overall, the ABM model can simulate individual citizens carrying out their daily activities. From this base activity, traffic within the city is simulated. SUMO's programmatic interface (named TraCI) is used to control the external simulation (e.g.: inject traffic flows, change traffic light status) and to access to the simulated data (e.g.: mean speed in streets, street occupancy…). Simultaneously, citizens may require medical care in a network of hospitals which is simulated based on the model already proposed in [41]. Within the model, it is possible to simulate the transmission of infective diseases based on the individual contacts of citizens as they visit different simulated places in their daily activities. The correct operation of both hospitals and traffic infrastructure requires the existence of ICT infrastructures (e.g.: servers, IT services…). The availability of such entities is also modelled, while being subjected to different cyberattacks, following the model proposed by these authors in [42]. Interdependences between the ICT system and the mobility and healthcare systems are partially considered. First, a cyberattack on the hospital's cyber-infrastructure might reduce the capacity and care quality of such hospital. On the other hand, a cyberattack on the cyber-infrastructure controlling traffic-lights might render them unavailable and affect traffic. Finally, the urban landscape layer provides geographical and physical context to the previous systems.

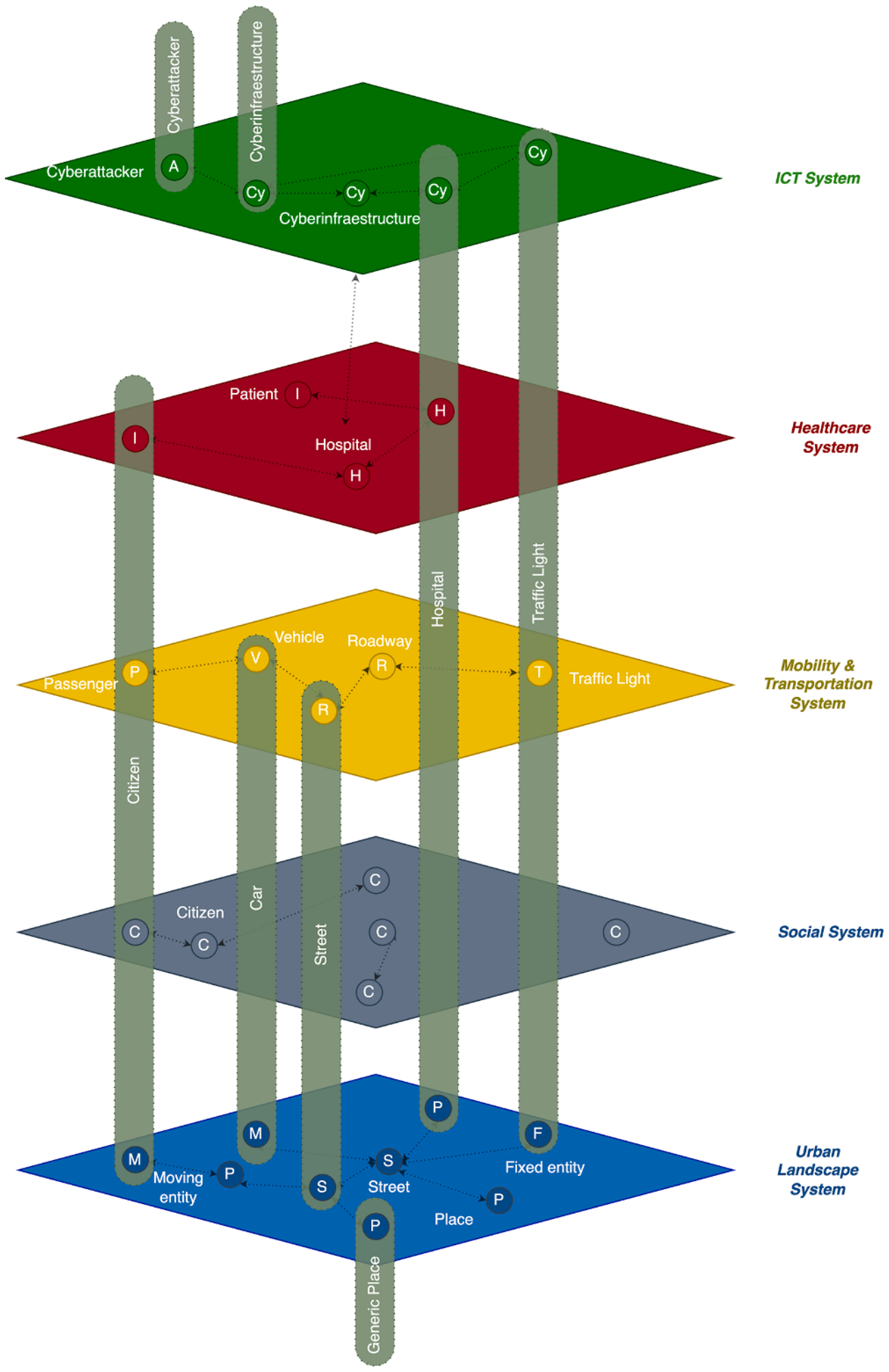

**Fig. 7.** Systems, agents, and sub-agents considered in the proof-of-concept of the ABM simulator.

Following the formal model, the city agents described in Table 4 are implemented in the simulator. To do so, the OOP methodology described in Section 4.2 is used. In it, each agent is derived from a set of individual sub-agents, each participating in a single system. For each subagent, a state and a series of internal and system rules have been defined and implemented, as briefly indicated in Table 5. A detailed explanation of each simulation model is out-of-the-scope of this paper, and the reader is kindly referred to the authors previous unintegrated work [41,42] for further details. Then, the already described interdependencies between systems are considered at agent-level, coordinating the behavior of the different sub-agents.

Finally, to introduce the simulation of the identified risks within the simulation, both the occurrence of cyberattacks and the spread of infective diseases can be manually scheduled. The list of contingencies could be easily extended if required to model other risks. Moreover, as the simulator is configurable with scenarios, the general framework described in this section can be particularized for the city description of Section 5.1 and to introduce mitigation measures by tuning the parameters of each agent.

Regarding the metrics, the simulator can compute the following system level metrics for each infrastructure-related systems based on the observable variables of each sub-agent (SA function):

- Mobility and transportation system SL: the mean relative speed reduction between the baseline scenario and the risk scenario for all roadways where a measurement station is present.

**Table 4**
Implementation details of each agent.

| Agent | Sub-agents | Description / AR |
|---|---|---|
| Cyber-attacker | Cyber-attacker | An entity that can generate botnet, DDOS and ransomware attacks on cyber-infrastructures. Attacks are propagated through a network of cyber-infrastructures respecting its topology. No direct dependencies are considered with other systems. |
| Cyber-infrastructure | Cyber-infrastructure | An ICT service provider that does not provide direct service to a hospital or traffic light. Cyber-infrastructures relate to each other forming a network of dependences. If a given infrastructure becomes unavailable due to an attack, this unavailability propagates to other dependent infrastructures. |
| Citizen | Citizen<br>Patient<br>Passenger<br>Moving entity | A citizen of the city. The citizen carries out its daily activities (citizen sub-agent) visiting different places (moving entity sub-agent) where it contacts with other citizens. It can use a vehicle to perform a given route between places (passenger sub-agent). As a result of the movement, a given infective disease can propagate (patient sub-agent) and require medical attention. |
| Hospital | Hospital<br>Cyber-infrastructure<br>Place | A hospital of the city (place sub-agent) that attends patients. Hospitals form a network (hospital sub-agents) referring patients between them to balance capacity. They depend on the correct operation of ICT infrastructure (cyber-infrastructure sub-agent) that can reduce the hospital capacity and care quality if unavailable. |
| Cars | Vehicle<br>Moving entity | A car moving (moving entity sub-agent) through the city with a given route (vehicle sub-agent) |
| Streets | Roadway<br>Street | A street of the city (street sub-agent) that is open to vehicles traffic (roadway sub-agent) |
| Generic place | Place | A generic facility/site/commerce of the city that can be visited by citizens during their daily activities. |
| Traffic Light | Fixed Entity<br>Traffic Light<br>Cyber-infrastructure | A traffic light governing a city intersection of the city (fixed entity sub-agent). It controls the flow of traffic (traffic-light sub-agent) as governed by its control element (cyber-infrastructure). In case of cyberattack, the traffic light might stop to operate (i.e.: turn off). |

Sub-agents column show the sub-agents included in each agent. Different colours are used to represent the system they belong to according to the color code already applied in Fig. 7. Description / AR column provides a description on the agent's behavior and the relationship between the different sub-agents.

$$SL_{mobility} = \frac{1}{N_{stations}} \sum_{\forall\ station} \min\left(\frac{speed^{risk}_{station}}{speed^{baseline}_{station}}, 1\right) \tag{11}$$

- Healthcare system SL: the mean number of unattended patients relative to the hospital capacity across all hospitals.

$$SL_{healthcare} = \frac{1}{N_{hospitals}} \sum_{\forall hospital} \max\left(0, 1 - \frac{\text{unattended patients}_{hospital}}{\text{capacity}_{hospital}}\right) \tag{12}$$

- ICT system SL: the percentage of available cyber-infrastructures over the total of cyber-infrastructures.

$$SL_{ICT} = \frac{n_{available}}{N_{cyberinfrastructures}} \tag{13}$$

In addition, at a city level, the service also computes the aggregated number of deaths due to any cause to assess the impact of risks on human casualties. These metrics will enable the data-driven risk-analysis and decision-making in next stages.

### 5.3. Stage 3. Risk analysis

Continuing with the third step of the methodology, the considered hazardous event (i.e.: infective disease + cyberattack) is parameterized within the simulator. Then, in order to evaluate the impact, two simulation scenarios are run: a baseline scenario and a scenario where the threats are experienced; and their results compared.

The decision maker can start by assessing the effect at system level to get an overall view of the impacts. As seen in Fig. 8, the main impact of the compounding threat is experienced in the healthcare system as its SL is dramatically decreased during most of the simulation. This reduction might be caused by the pandemic extension through the city that increases the healthcare demand. At the same time, reductions in the SL of the ICT and Mobility systems are also observed but in a more contained way, both in terms of duration and SL reduction. This limited impact could be explained by the cyberattack whose effect propagates to other systems. In fact, the healthcare system seems to also be affected as demonstrated by the first, reduced drop in service level concurrent with the other SL drops.

**Table 5**
Implementation details of each sub-agent.

| Sub-Agent | Description | Parameters | State | IR | SR | OB |
|---|---|---|---|---|---|---|
| Cyber-attacker | Representation of any entity carrying out a cyber-attack. | - Target<br>- Type of attack (botnet, ransomware, DDOS) | - Attack status | - Attack generation | - Attack transmission | - Number of attacked entities |
| Cyber-infrastructure | Representation of an ICT service provider or physical server. | - Service Capacity<br>- Vulnerability level<br>- Recovery Speed<br>- Network Topology | - Availability status | - Nominal operation.<br>recovery logic | - Dependence on other cyber-infrastructures.<br>- Attack defense and re-transmission logic | - Availability status |
| Patient | Representation of the health situation of any citizen. | - Vaccination status.<br>- Initial infective status. | -Infective status<br>Health status | - Infection and disease processes.<br>-Base medical needs. | -Disease transmission<br>Seeking medical attention in hospital | -Infection status |
| Hospital | Representation of a hospital. | - General beds capacity<br>- ICU beds capacity<br>- Healthcare quality<br>- Healthcare system topology | -General bed occupancy<br>ICU bed occupancy | - Medical care process (recovery process of patients) | -Patient reception /discharge<br>- Patient referral to other hospitals | -General bed occupancy<br>- ICU bed occupancy<br>- Number of unattended patients |
| Passenger | Representation of a traveler to interact with the SUMO simulator | N/A | - Position<br>- Route | N/A | - Inserts route into SUMO simulator | N/A |
| Vehicle | Representation of a vehicle from the SUMO simulator | - Managed by SUMO external simulator | - Position<br>- Speed<br>- Route | - Managed by SUMO external simulator | - Managed by SUMO external simulator | N/A |
| Roadway | Representation of a street from the SUMO simulator | - Measurement station existence<br>- Max speed | - Managed by SUMO external simulator | - Managed by SUMO external simulator | - Managed by SUMO external simulator | - Mean speed<br>- Traffic intensity |
| Traffic Light | Representation of a traffic light from the SUMO simulator. | - Managed by SUMO external simulator | - Operation status | - Managed by SUMO external simulator | - Managed by SUMO external simulator | - Operation status |
| Citizen | Representation of a citizen of the city carrying out its daily activities following a statistical pattern. | -Social timetable<br>- Social graph | - Current activity | -Daily activities evolution | -Contact generation. | -Current activity |
| Moving Entity | Representation of any entity that moves over the urban landscape | -Initial position | - Position<br>-Current place<br>- Speed | N/A | N/A | -Current place |
| Place | Representation of any facility/site/commerce that can be visited by citizens. | - Position<br>-Maximum capacity | - Occupancy | N/A | N/A | - Occupancy<br>- Occupants |
| Street | Representation of a street. | - Position<br>- Capacity<br>-Linked streets | N/A | N/A | N/A | N/A |
| Fixed Entity | Representation of any fixed object/entity present in the urban landscape. | - Position | N/A | N/A | N/A | N/A |

In the sub-agent column, different colours are used to represent the system they belong to according to the color code already applied in Fig. 7. A description of each sub-agent is provided along with the main parameters, state, functionality, and observable variables.

Another analysis of interest can be performed by studying the level of service with a geographical component. In this case, the effects of the cyber-attack are not limited to the attacked district but also spread, albeit to a lesser extent, to the other district (as shown in the cases of healthcare and mobility). On the other hand, the effects of the pandemic are more symmetrical in both districts.

The analysis above, based on high-level metrics, is preliminary and allows the user to get a general idea of the main affected systems. Intuitions can be confirmed and extended by consulting the specific metrics of each agent, which is possible thanks to the granularity of ABM simulation. In this sense, Fig. 9 depicts the direct effects of the simulated threats: the disease propagates exponentially following the typical curve of SIR models and the cyberattack affects to all infrastructures within the center district but not to the outskirts district.

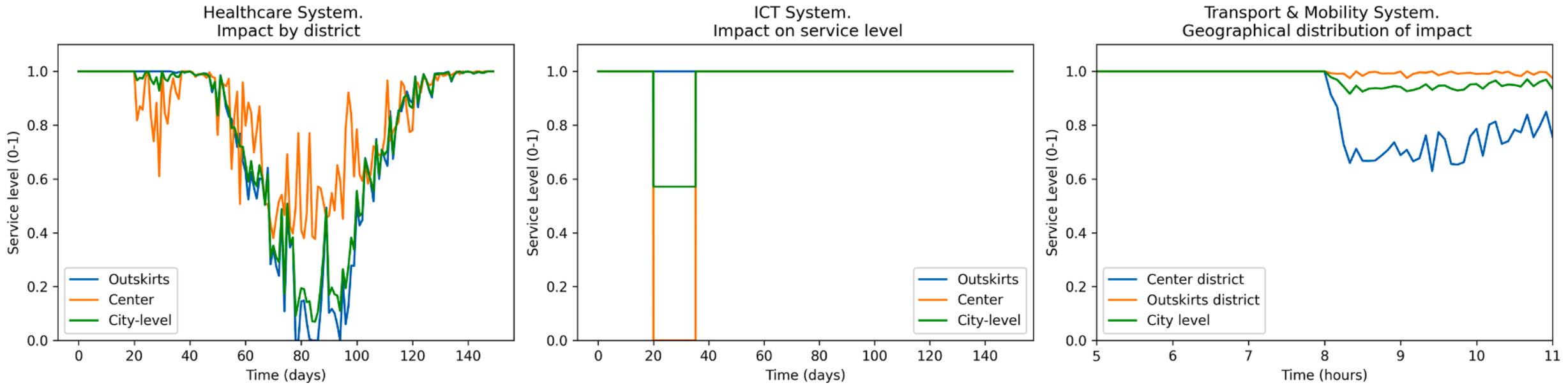

**Fig. 8.** Impact at system level.
Service level of the healthcare system (left), ICT system (center) and transport and mobility system (right, a zoomed time scale centered around day 20 is used here to improve the visualization).

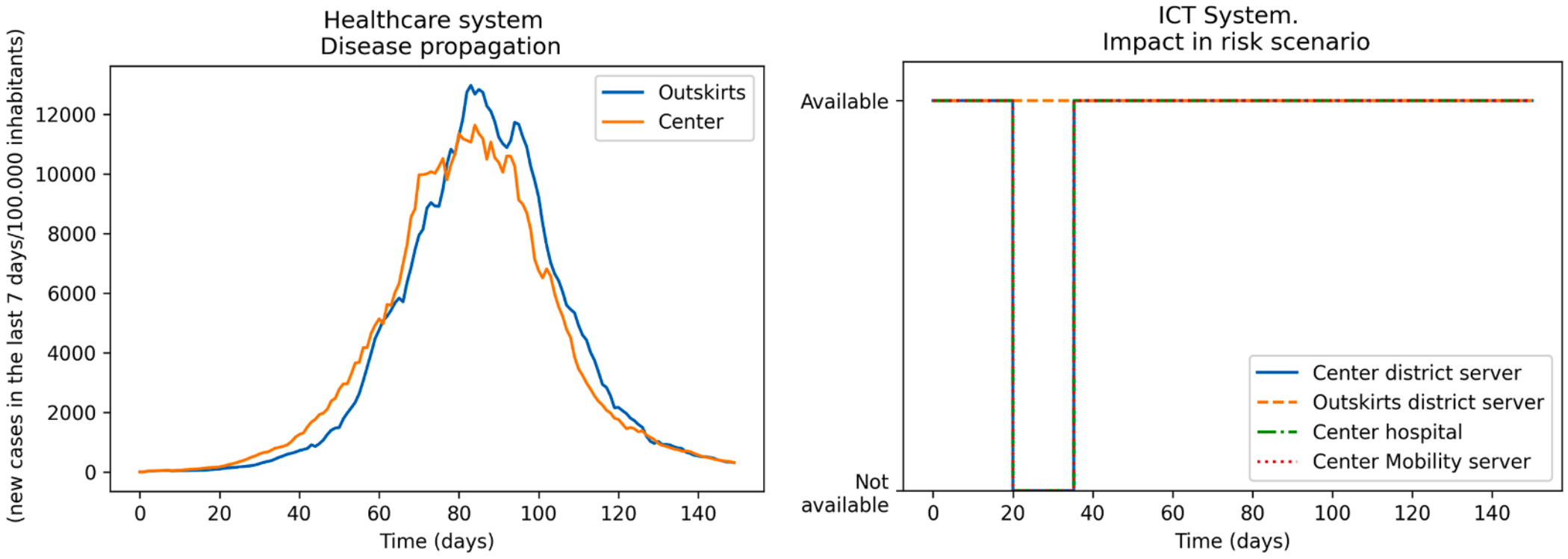

**Fig. 9.** Threat simulation results.
Infective disease propagation (left). Cyberattack occurrence (right).

Fig. 10, below, shows the effect on the two simulated hospitals in terms of their individual sub-agent level metrics. Cross-checking these results with the previous figures allows to conclude that the first drop in the SL of the center district is due to a reduction in bed capacity in its hospital caused by the cyberattack. At the same time, it can be shown that this generates an increase in patients at the outskirts hospital as patients are transferred there. Continuing with the simultaneous analysis of both graphs, the second significant SL drop corresponds to the increase in the number of people infected and the total occupancy of the two hospitals. In terms of mobility, it is also confirmed that the occurrence of the cyber-attack correlates temporarily with the change in the speed and occupancy parameters of the downtown streets. On the other hand, the streets in the outskirts are less affected as their traffic lights continue to operate.

Summing up, **the proposed simulation framework can serve as a DSS tool that allows the quantitative analysis of the impact of threats at different levels. First, system-level metrics can be used to get a broad view on the city resilience. Then, agent-level metrics back those figures and provide better explainability**.

### *5.4. Steps 4 & 5. Evaluating mitigation alternatives and taking a risk-informed decision*

This previous risk analysis process can serve as the basis of risk-based decision-making (steps 4 and 5 of the methodology). Evaluating the impact figures, the policymaker recognizes that the impact on the city systems is not acceptable. To cope with the risk and enhance the city resiliency, he proposes two alternative mitigation solutions, limited by the available resources. The previous results are used to generate sensible alternatives that tackle the major affectations in the city. In that sense, it has been detected that the health system is strongly degraded and that the failures in the ICT infrastructure can spread throughout the city. Therefore, the proposed alternatives to alleviate these problems are:

- Mitigation alternative 1: beds. It consists in acquiring a quickly deployable capability that increases by 50% the capacity of each hospital. The measure intends to avoid hospital overload, reducing the number of unattended patients.
- Mitigation alternative 2: cybersecurity. It consists in improving the cybersecurity level of the city both in terms of countermeasures and incident response times. It tries to avoid cyber-infrastructure downtime whose effects propagate to other interdependent systems.

Each of the alternatives is parameterized within the simulator and simulations are run to assess the impact reduction in each case.

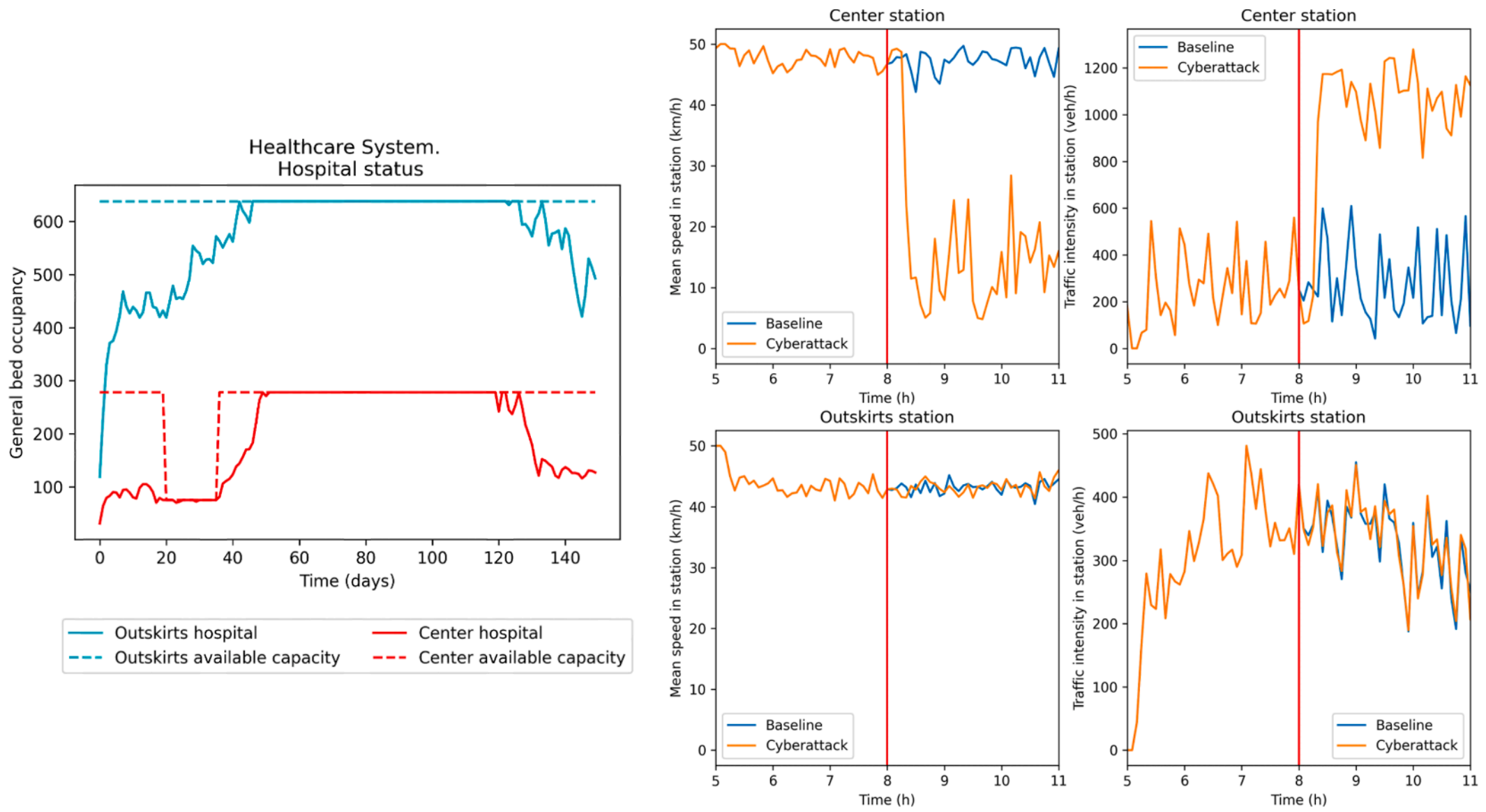

**Fig. 10.** Impact of the threat at agent-level.
Hospital occupancy and available capacity (left). Traffic in a measurement station of the center district (right, top) and the outskirts district (right, down).

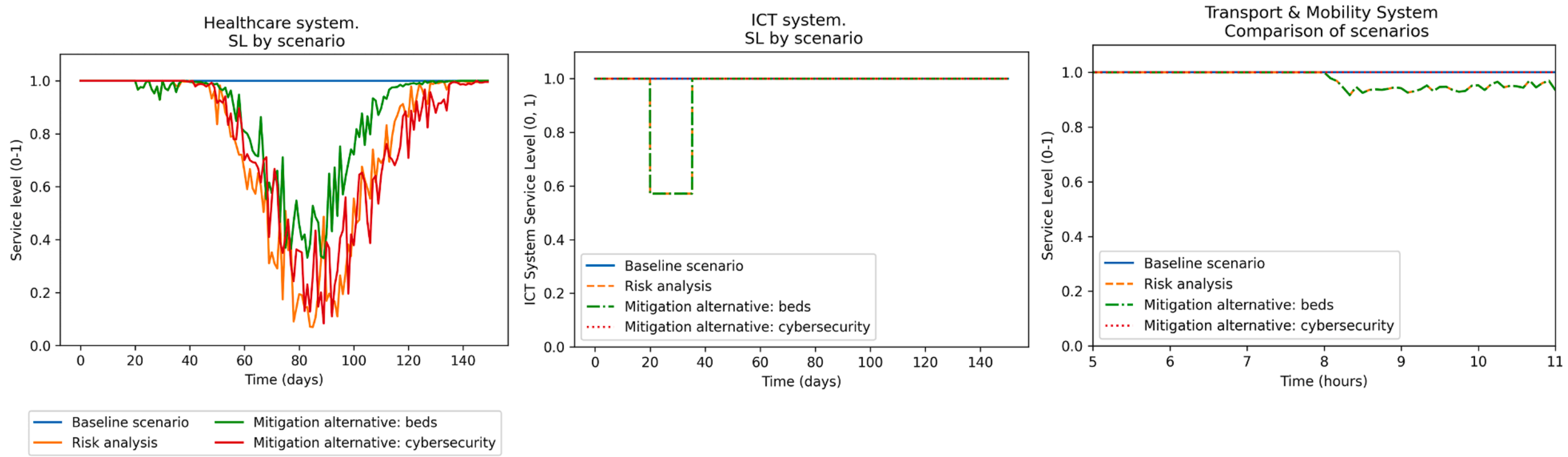

**Fig. 11.** Risk reduction at system level.
Effect on service level of the healthcare system (left), ICT system (center) and transport and mobility system (right, a zoomed time scale centered around day 20 is used here to improve the visualization).

In this sense, Fig. 11 compares the SL of each system for the four considered scenarios in the validation use case: baseline scenario, risk analysis, mitigation alternative 1 and 2. It can be shown that, as it could be expected, mitigation alternative 1 does not yield an improvement in the service level of the ICT and mobility systems. However, the metric of the healthcare system is significantly improved even if hospital overload is not completely avoided. In turn, the second mitigation alternative eliminates the cyber-attack effect on the ICT system and the cascading effects on the healthcare and mobility systems. Nevertheless, the healthcare system server level hardly improves when compared with the baseline case.

In cases where analysis by system is not clearly conclusive, one can resort to the aggregated city-level metrics also provided by the framework. In this regard, Fig. 12 shows the cumulative number of deaths citywide for each scenario. Alternatively, a scoring system could also have been used with the SLs, although this would entail prioritizing some systems over others. Focusing on the effects on deaths, it seems clear that the best alternative is to increase hospital capacity. However, the impact is still high, which could lead the decision-maker to generate new alternatives in an iterative process in order to maximize the impact reduction.

The previous use case was used to demonstrate the usefulness of the proposed simulation framework as a **data-driven DSS tool for comparing decision alternatives based on a reliable qualitative risk assessment of each one of them**. The inclusion of this type of tools in the urban planning decision-making process can help in improving the resilience of cities.

## 6. Conclusions

As the importance of cities increases in modern societies, so does the impact of the occurrence of threats and hazards. To guarantee the safety of city inhabitants and the normal operation of the urban space, the concept of resilience and risk management must be introduced into the urban planning practices. However, due to the complexity of cities, risk analysis and assessment is challenging and often only partially carried out without considering the interdependencies among city domains. Among the article's main contributions, in first place, a key element is the **comprehensive conceptual model of the city**, which aims at overcoming the domain-specific approaches and explores the city as a set of interrelated systems linked by a Social System deployed over the Urban Landscape. Contrary to other theoretical examples in literature, the contribution aims to provide a general view of the city covering all dependencies.

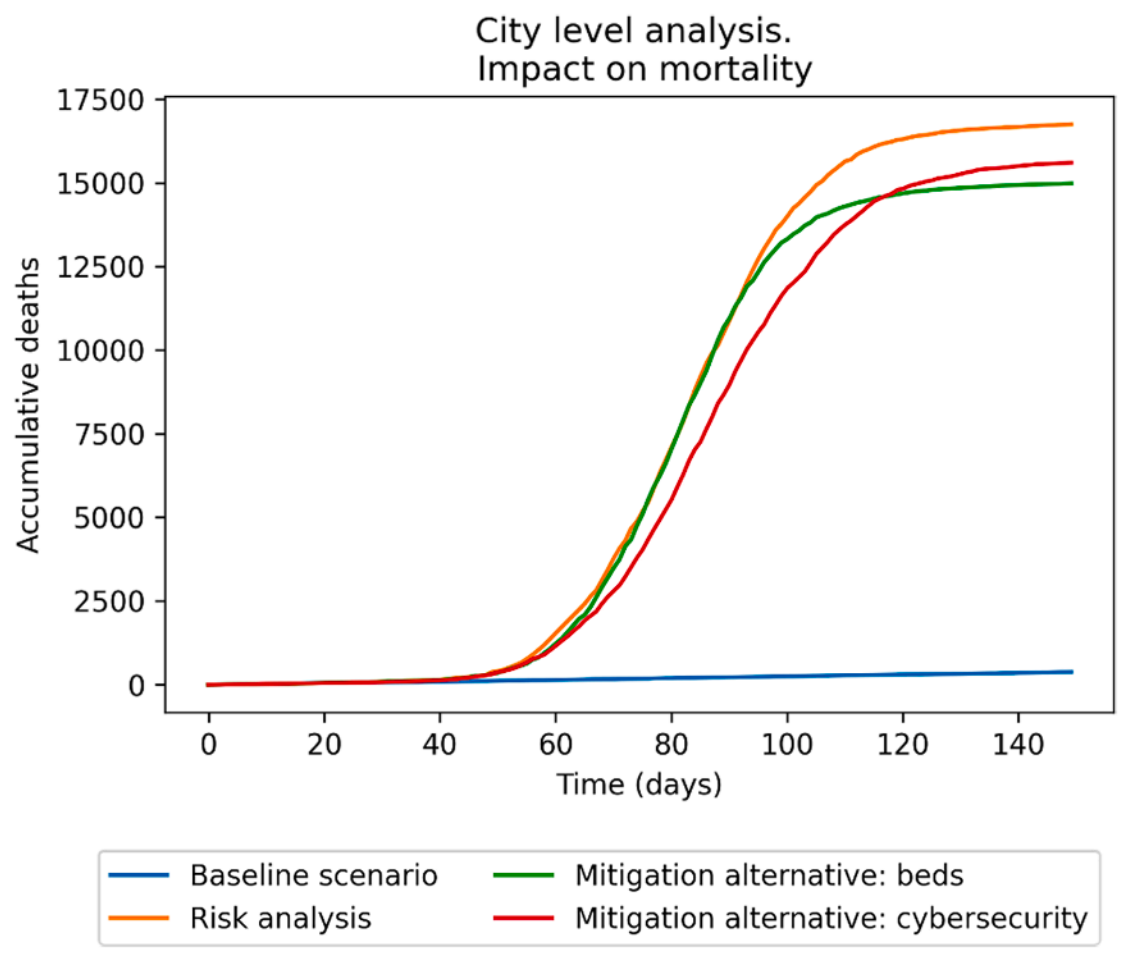

**Fig. 12.** Risk reduction at city-level. Effect on observed mortality.

As a possible solution to better manage uncertainty in cities, this article has proposed the use of a comprehensive, **hybrid simulation framework** as a decision support system to guide risk analysis with quantitative data. Hybrid methods emerge as superior choices for urban city modeling compared to simpler approaches. While empirical techniques and agent-based modeling have limitations in estimating new disasters and validation complexities, respectively, system dynamics methods face challenges in analysing component-level behavior. In turn, economic theory approaches may not cover all systems and relationships, and stand-alone network-based methods struggle with representing isolated CI complexity. In contrast, hybrid-based methods integrate various modeling approaches, achieving a balance between micro and macro dynamics. Despite an increase in design model complexity, they offer a comprehensive and nuanced representation of urban dynamics, making them more adept for city modeling in diverse and complex environments.

In this article, the proposed practical design of a hybrid simulation framework takes advantage of the benefits of ABM and network-based modeling. Agent-based modeling (ABM) offers a method for investigating the emergent behaviours within a city, while network-based modeling enables the seamless representation of interdependencies among various entities. As a result, the city is conceived as a collection of independent agents, each comprised of subagents and incorporating stochastic elements. This framework facilitates the creation of functional simulation models that offer insights into city trends. Additionally, as differentiating contribution, it organizes entity dependencies in a manner conducive to easy computer implementation, enables simulator federation and discrete simulation in accelerated time. A key aspect is that the framework mirrors the smart-city paradigm by providing multi-level simulated metrics on city operation, thereby facilitating decision-making-oriented visualization.

The use of the framework implies the application of a well-defined five-stage methodology that starts with problem scope definition (identifying relevant risks and city systems to consider, stage 1). Then, a modeler generates an actionable simulation model that it deployed on the hybrid simulation framework (stage 2) to enable data-driven risk analysis using metrics at various aggregation levels (stage 3). Risks are subsequently evaluated against acceptable thresholds, and mitigation alternatives are proposed and assessed for their effectiveness in risk reduction using the simulation mode (stage 4), before using the insights gained to make a final risk-informed decision aimed at enhancing city resiliency (stage 5). While the city modeling effort persists, the presence of a clearly articulated model with layered metrics, guiding system design, and the adaptability of the implementation approach collectively mitigate the complexity of the work.

To support this final statement, the **simulation framework has been applied for a novel subset of systems ICT system, mobility system and healthcare system** used to deploy a demonstration scenario. The implementation exemplifies how the tailored framework can work as a data-driven DSS tool that enables risk-informed decision making, by assessing different risk scenarios and decision alternatives. Although the final implementation of the DSS and simulation used for validation has been limited to three systems due to the time-consuming nature of the modeling process, the described city concept and simulation framework are general thus they can be easily extended.

Future work could involve incorporating external relationships to better reflect the interconnected reality of cities. The current model treats the city as a self-contained entity, but the modeling offers potential to enrich it by considering relationships with other cities, national structures, and relevant external entities through the "boundary entities" concept already considered in the model. Additional lines of research also include the exploration of artificial intelligence (AI) for outcomes' explainability, enabling a deeper understanding of simulation results for decision making. How to activate a semi-automated workflow that is triggered by anomaly detection and focuses on providing prescriptive analysis may also derive into an innovative proposal. And looking ahead, the integration of systems for fully automated decision making within self-adaptive urban simulations represents an evolving frontier, presenting opportunities to enhance the efficiency and responsiveness of city models.

## Data availability

No data was used for the research described in the article.

## Acknowledgements

Authors acknowledge the support received under grant IA4TES (funded by the Spanish Ministry of Economic Affairs and Digital Transformation) and by Project RP220022063 (Universidad Politécnica de Madrid).

## Annex 1. Influence diagram representing interdependences among city systems

The following figure provides a comprehensive view of the interdependencies among different systems. It has been obtained by aggregating a systematic analysis that identifies system by system their main interdependences with other systems. The outgoing arrows from each system indicate that the source system provides a service or supplies a product to the target system as indicated in the arrow tag:

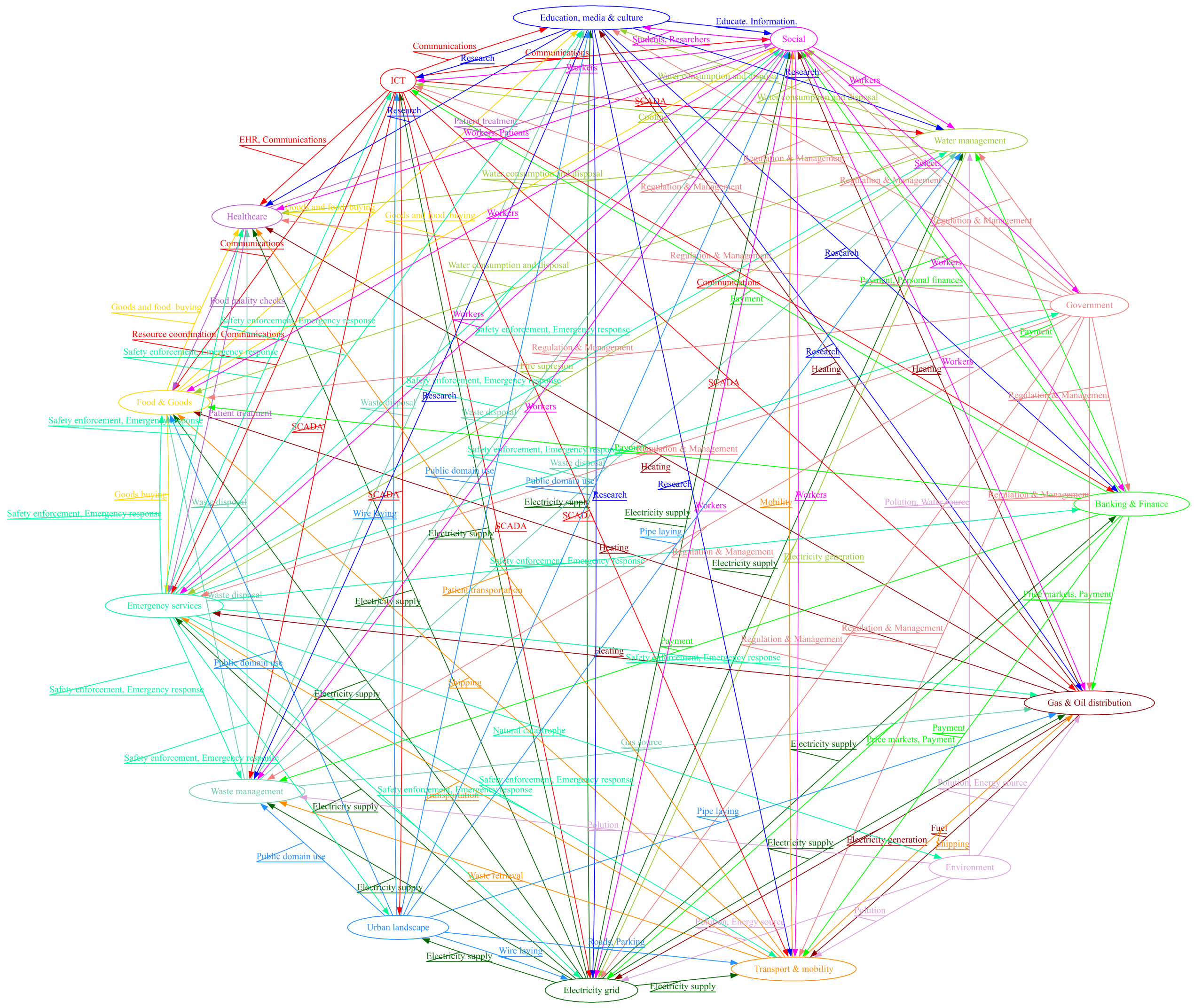

## Glossary

*ABM:* Agent Based Model/Modelling
*AOB:* Aggregated observability Function
*AR:* Agent interaction Rules
*CI:* Critical Infrastructure
*CAS:* Complex Adaptative Systems
*DSS:* Decision Support System
*HE:* Hazardous Event
*IR:* subagent Internal Rules
*OB:* Observability rules
*OOP:* Object Oriented Programming
*SA:* System Aggregation
*SL:* Service Level
*SR:* System interaction Rules